\documentclass[12pt,preprint]{aastex}

\begin{document}

\shorttitle{Partial Ionization in the Chromosphere}
\shortauthors{Mart\'inez-Sykora \& De Pontieu \& Hansteen }

\title{2D Radiative MHD Simulations of the Importance of Partial Ionization in the Chromosphere}

\author{Juan Mart\'inez-Sykora $^{1,2}$}
\email{j.m.sykora@astro.uio.no}
\author{Bart De Pontieu $^{1}$}
\and
\author{Viggo Hansteen $^{1,2}$}

\affil{$^1$ Lockheed Martin Solar and Astrophysics Laboratory, Palo Alto, CA 94304}
\affil{$^2$ Institute of Theoretical Astrophysics, University of Oslo, P.O. Box 1029 Blindern, N-0315 Oslo, Norway}

\newcommand{\myemail}{juanms@astro.uio.no}
\newcommand{\viscous}{\underline{\underline{\tau}}}
\newcommand{\resistive}{\underline{\underline{\eta}}}
\newcommand{\komment}[1]{\texttt{#1}}

\begin{abstract}

The bulk of the solar chromosphere is weakly ionized and interactions between 
ionized particles and neutral particles likely have significant consequences for the 
thermodynamics of the chromospheric plasma. We investigate the importance 
of introducing neutral particles into the MHD equations using numerical 2.5D 
radiative MHD simulations obtained with the Bifrost code. The models span 
the solar atmosphere from the upper layers of the convection zone to the low corona, 
and solve the full MHD equations with non-grey and non-LTE radiative transfer, 
and thermal conduction along the magnetic field. The effects of partial ionization 
are implemented using the generalized Ohm's law, i.e., we consider the effects 
of the Hall term and ambipolar diffusion in the induction equation. The approximations
required in going from three fluids to the generalized Ohm's law are tested in our 
simulations. The Ohmic diffusion, the Hall term, and ambipolar diffusion show strong variations in 
the chromosphere. These strong variations of the various magnetic diffusivities 
are absent or significantly underestimated when, as has been common
for these types of studies, using the semi-empirical 
VAL-C model as a basis for estimates. In addition, we find that differences in estimating
the magnitude of ambipolar diffusion arise depending on which method is used to 
calculate the ion-neutral collision frequency. These differences cause uncertainties
in the different magnetic diffusivity terms.  In the chromosphere, we find that the 
ambipolar diffusion is of the same order of magnitude or even larger than the 
numerical diffusion used to stabilize our code. As a consequence,
ambipolar diffusion produces a strong impact on the modeled atmosphere. 
Perhaps more importantly, it suggests that at least in the chromospheric domain, 
self-consistent simulations of the solar atmosphere driven by magneto-convection 
can accurately describe the impact of the dominant form of resistivity, i.e., 
ambipolar diffusion. This suggests that such simulations may be more
realistic in their approach to the lower solar atmosphere (which
directly drives the coronal volume) than previously assumed.
\end{abstract}


\section{Introduction}

Most of the models and simulations of the solar atmosphere solve the 
magnetohydrodynamics (MHD) equations, implicitly assuming that the 
plasma is magnetized, i.e., fully ionized or with the ion-neutral collision 
frequency lower than the ion gyrofrequency \citep[][among others]
{Schaffenberger:2005fj,Vogler:2005fj,Stein:2006qy,Gudiksen:2011qy}. 
However, since the photosphere and parts of the chromosphere are 
unmagnetized, i.e., the ions are not necessarily tied to the field lines, we 
expect that neutral particles can have a significant impact on the dynamics of this region 
\citep{Vernazza:1981yq,Fontenla:1990yq,Fontenla:1993fj}. Therefore, 
it is likely that, under some conditions, the photosphere and chromosphere 
should be treated as a three component fluid, where the dynamics of the 
neutrals, ions, and electrons are treated separately. Under the assumption 
of a weakly ionized plasma, one can return to a one-component fluid. However, 
new terms known as the Hall term and ambipolar diffusion appear in the induction equation 
\citep{Parker:1963vn,Parker:2007lr,Pandey:2008qy}. The latter term is a consequence 
of the ion-neutral dissipation which can be derived from the Cowling resistivity
\citep{Khodachenko:2004vn,Khodachenko:2006kx,Leake:2006kx}. 
This form of the induction equation is known as the generalized Ohm's law 
\citep{cowling1957}. 

A large number of papers in recent years have investigated the effects of the 
ion-neutral interactions on single fluid MHD. \citet{Leake:2006kx} simulated 
2.5D simulations of flux emergence and observed that the ambipolar diffusion 
leads to an increase of the rates of magnetic field emergence and a
resultant magnetic field that is much more diffuse than the case with
only Ohmic diffusivity. In addition, the magnetic field that emerges into the corona 
is found to be more force-free, since currents are aligned to the
field. This is because ambipolar diffusion acts on the currents
perpendicular to the magnetic field. \citet{Arber:2007yf} extended this simulation to 3D where the 
previous results were confirmed, and in addition found that, as a result of
including neutrals, flux emergence lifts less chromospheric material
to great heights. This effect suppresses the Rayleigh-Taylor instability between the 
emerging flux and the corona. 

The interaction between ions and neutrals can also dissipate Alfv\'en waves as 
result of the small but finite coupling time between ions and neutrals. This type of damping 
can heat and accelerate the plasma in the upper chromosphere and in spicules 
\citep{de-Pontieu:1998lr,de-Pontieu:1999uq,James:2002lr,James:2003fk,Erdelyi:2004qy}
and incur wave energy leakage at the footpoints of coronal loops \citep{De-Pontieu:2001fj}
and in the network
\citep{Goodman:2000ys}. \citet{Khodachenko:2004vn}, using the
temperature and density structure from the 1D VAL-C model, 
concluded that the collisional friction damping of MHD waves is often more 
important than the viscous damping for waves propagating in the partially ionized 
plasmas of the solar photosphere, chromosphere and prominences. Estimates of the 
efficiency of the damping of waves were made by \citet{Leake:2005rt}
as well as by the previous authors. 

\citet{Pandey:2008qy} determine that waves can be affected by the Hall
term at both low and high fractional ionization, because the Hall regime 
wave damping is inversely proportional to the fractional ionization. Thus 
Hall term may also be important at high fractional ionization in
contrast to ambipolar diffusion which is important only at low fractional ionization. 

\citet{Khomenko:2012ys} performed various simplified scenarios where they studied the
impact of the ambipolar diffusion in the chromosphere. They conclude that current 
dissipation enhanced by the action of ambipolar diffusion is an important process that 
is able to provide a significant energy input into the chromosphere. Heating from 
ambipolar diffusion leads to thermodynamic evolution in the chromosphere on
timescales of about 10--100 seconds.

All the models above, even the 2D and 3D models, are based on a 1D
semi-empirical atmosphere (e.g., VAL-C), and/or a simplified approach to the 
energy balance in the chromosphere (adiabatic or Newtonian cooling). In 
addition, none of the partial ionization effects have been considered in 
full magneto-convection simulations. \citet{Cheung:2012vn} have made progress
in this direction and performed full magneto-convection simulations of an umbra
taking into account partial-ionization effects. However, their simulations
only extend up to the upper photosphere. 

In this paper, we use the Bifrost code 
\citep{Gudiksen:2011qy} to create a self-consistent and fully dynamic model 
atmosphere of the sun, from the convection zone to the corona, to consider 
the importance of the Hall term and ambipolar diffusion relative to the Ohmic and 
artificial diffusion. Unlike other models, Bifrost includes an advanced treatment of
radiative losses in the chromosphere based on recipes derived from
dynamic non-LTE radiative 1D hydrodynamic simulations. Such a treatment
is crucial for a consideration of the effects of partial ionization, as shown in 
what follows. The code and the implementation of the generalized Ohm's law 
are described in Section~\ref{sec:equations}. The tests performed for the code 
validation are discussed in Section~\ref{sec:test}. We describe the different 
forms of diffusion in 2D MHD simulations in Section~\ref{sec:diff}. Finally, the 
various simplifications made in order to obtain the generalized Ohm's law 
following \citet{Pandey:2008qy} have been investigated 
and tested for the 2D MHD simulations in Section~\ref{sec:validation}. 
The paper finishes by addressing the conclusions and discussion. 

\section{Equations and numerical method}
\label{sec:equations}

The magnetic upper-photosphere and chromosphere is weakly ionized
and the interaction between ionized particles and neutral particles 
potentially has important consequences for the thermodynamics 
\citep{Fontenla:1993fj} of this region.
We investigate these consequences in the solar atmosphere. In order to 
model the solar atmosphere we solve the MHD equations in $2.5$D.
The model spans from the upper layers of the convection zone to the low 
corona. We have implemented the effects of partial ionization into the
induction equation through the Hall and ambipolar 
diffusion terms as described below.

The Bifrost \citep{Gudiksen:2011qy} code is a staggered mesh,
explicit code that solves the MHD partial differential equations,
including non-LTE and non-grey radiative transfer with scattering, 
and conduction along the magnetic field lines. A lookup table, based on
LTE, is used to compute the temperature, pressure, opacities and other
radiation quantities, and
ionization state, given the pressure and the internal energy of the
plasma. Spatial derivatives and the interpolation of variables are done
using high order polynomials. The equations are stepped forward in time using the explicit 
third order predictor-corrector procedure described by \citet{Hyman1979}.
In order to suppress numerical noise, high-order artificial diffusion is added 
both in the forms of a viscosity and in the form of a magnetic diffusivity 
\citep[see][for details]{Gudiksen:2011qy}. 

The Bifrost code includes an advanced treatment of the effects of
radiation on the local energy balance, which is crucial if one wants
to accurately determine the ionization degree. 
The radiative flux divergence from the photosphere and lower chromosphere 
is obtained by angle and wavelength integration of the transport equation 
assuming isotropic opacities and emissivities. The transport equation 
assumes that opacities are in LTE using four group mean opacities to cover
the entire spectrum \citep{Nordlund1982}. This is done by formulating the 
transfer equation for each of the four bins, calculating a 
mean source function in each bin. These source functions contain an 
approximate coherent scattering term and an exact contribution from thermal
emissivity. The resulting 3D scattering problems are solved by iteration, 
based on one-ray approximation in the angle integral for the mean intensity,
a method developed by \citet{Skartlien2000}.

In the mid and upper chromosphere, the Bifrost code includes non-LTE radiative 
losses from tabulated hydrogen continua, hydrogen lines, and lines from singly 
ionized calcium as functions of temperature and column mass \citep{Carlsson:2012uq}. These radiative 
losses depend on the computed non-LTE escape probability as a function 
of column mass and are based on a 1D dynamical chromospheric model in which 
the radiative losses are computed in detail 
\citep{Carlsson:1992kl,Carlsson+Stein1994,Carlsson:1997tg,Carlsson:2002wl}.

The energy dissipated by Joule heating is given by $Q_{Joule}={\bf E \cdot J}$  
where the electric field ${\bf E}$ is calculated from the current ${\bf J}$,
taking into account high-order artificial resistivity. The resistivity
is computed using a hyper-diffusion operator \citep{Gudiksen:2011qy}. 
This entails that the Joule heating due artificial diffusion 
is set proportional to the current squared times a factor that becomes
large (of order 10) when magnetic field gradients are large, and is unity otherwise.

\subsection{Generalized Ohm's Law theory}
\subsubsection{Multi-fluid}

Most codes treat the solar atmosphere as a single fluid where collisional frequencies 
are considered sufficient to ensure that all species are well coupled and that the 
momentum and energy equations can be added without the introduction of frictional 
terms or similar. However, as chromospheric temperatures are likely to drop to a few 
$10^3$~K or even lower \citep{Leenaarts:2011qy}, there is a high probability that plasma is
only partially ionized and that 
``slippage'' effects could become important. In this case the MHD equations should 
be treated by considering the plasma to consist of three fluids: ions, electrons and 
neutral particles. The mass density for each type of particle is governed by the continuity 
equation applied to each species separately: 

\begin{eqnarray}
\frac{\partial \rho_j}{\partial t} + \nabla \cdot \rho_j {\bf u_j} = 0\label{eq:contm}
\end{eqnarray}

\noindent where $\rho_j=m_j n_j$, ${\bf u_j}$, $n_j$ and $m_j$ are 
the mass density, velocity, number density and particle mass of the 
ion, electron, and neutral species, i.e., $j=\mathrm{i,e,n}$ respectively. 
The mass transfer term as result of ionization and recombination 
has been neglected. This approximation is valid for a one-fluid approach 
if the system is in ionization balance, and there is no decoupling of ions
and neutrals.

The momentum equation, written in SI units, for each species, is as follows:  

\begin{eqnarray}
\rho_i \left( \frac{\partial }{\partial t}+ {\bf u_i} \cdot \nabla \right) {\bf u_i} = - \nabla P_i + 
n_i Z q_e \left({\bf E} + {\bf u_i} \times {\bf B} \right) -\rho_i\sum_{j=e,n} \nu_{ij}({\bf u_i - u_j}) 
\label{eq:spmoma1}\\
\rho_e \left( \frac{\partial }{\partial t}+ {\bf u_e} \cdot \nabla \right) {\bf u_e} = - \nabla P_e - 
n_i q_e \left({\bf E} + {\bf u_e} \times {\bf B} \right) -\rho_e\sum_{j=i,n} \nu_{ej}({\bf u_e - u_j})
\label{eq:spmomb1}\\
\rho_n \left( \frac{\partial }{\partial t}+ {\bf u_n} \cdot \nabla \right) {\bf u_n} = - \nabla P_n- 
\rho_n\sum_{j=e,i} \nu_{nj}({\bf u_n - u_j})\label{eq:spmomc1}
\end{eqnarray}

\noindent where  $q_e$ and $Z$ are respectively the electron charge and ion 
charge. ${\bf E}$ and ${\bf B}$ are the electric and magnetic field and 
$P_j=n_{i}kT_{i}$ is the partial pressure of the $j$th species,  $k$ is Boltzmann's 
constant, and $\nu_{ij}$ is the collision frequency for species $i$ with species $j$. 
We assume that collisions are sufficiently numerous that the ion and electron 
temperature can be considered the same ($T_{i}=T_{e}$). All three equations 
are linked through the last term, i.e., the exchange of momentum between the
particles, where we have ignored the thermal force. In a similar manner as 
for the continuity equation, the momentum transfer term as result of 
ionization and recombination has been neglected.

The number of equations thus increases considerably compared with
single fluid MHD, but by considering some simplifications, as
described by \citet{cowling1957,Parker:2007lr,Pandey:2008qy}, one can 
easily generalize the MHD equations for each species to a single fluid (see below too). 
Therefore, the mass density is governed by the continuity equation for the bulk 
fluid as follows:

\begin{eqnarray}
\frac{\partial \rho}{\partial t} + \nabla \cdot \rho {\bf u} = 0 \label{eq:contb}
\end{eqnarray}

\noindent where  the density for the bulk fluid is the sum of the different particle 
densities ($\rho= \rho_i +\rho_e + \rho_n$), and considering 
$ \rho_i/\rho>> \rho_e/\rho$, then $\rho \approx  \rho_i +\rho_n$. 
In a similar manner, the velocity of the bulk fluid is 
${\bf u} \approx (\rho_i {\bf u_i}+\rho_n {\bf u_n})/\rho$, where the electron inertia is 
implicitly neglected in the definition of the bulk velocity. If we define the neutral 
density fraction ($D=\rho_n/\rho$), then ${\bf u} \approx (1-D){\bf u_i}+D {\bf u_n}$. 
Finally, the current density is given by ${\bf J}=n_e q_e({\bf
  u_i-u_e})$ (assuming singly charged ions). Since
Equation~\ref{eq:contb} is the same as for the single fluid formulation, the continuity 
equation does not need any modification in the Bifrost code.

Following \citet{Pandey:2008qy}, the single fluid momentum equation can be
recovered if we neglect the effects of the electron inertia. Because it is implicitly 
neglected in the definition of the bulk velocity, it can also be neglected 
in the continuity and momentum equations. For simplicity, the ions are
assumed to be singly charged, and we adopt charge neutrality ($n_{i}=n_{e}$). 
In addition, the drift momentum is assumed to be considerably smaller than 
the fast momentum ($\rho \sqrt{v_{a}^{2}+c_{s}^{2}}$) so that: 

\begin{eqnarray}
\rho_i\rho_n u_D^2 << \rho^2(v_a^2+c_s^2) \label{eq:neg1}, 
\end{eqnarray} 

\noindent where ${\bf u_{D}}={\bf u_{i}}-{\bf u_{n}}$, $v_a=B/\sqrt{ \mu_o \rho}$, 
and $c_s=\sqrt{\gamma P/\rho}$ are respectively the drift, Alfv\'en and sound 
velocities in the bulk fluid; $\mu_o$ is the vacuum permeability, and  
$\gamma$ is the ratio of specific heats. When the plasma does not fulfill  
Equation~\ref{eq:neg1}, the fluids are strongly decoupled. This happens
when the 
ion-neutral collision frequency is low.  When the drift momentum is low, the
drift momentum can be neglected for small dynamical frequencies (i.e.,
changes of the plasma properties on timescales commensurate with such frequencies):

\begin{eqnarray}
\omega \leq \frac{\rho}{\sqrt{\rho_i \rho_n}} \left( \frac{D\beta_e}{1+D\beta_e}\right) \nu_{ni} \label{eq:neg5} 
\end{eqnarray}

\noindent where $\beta_{\rm e}=\frac{\omega_{c\mathrm{e}}}{\nu_{\rm
    e}}$, the ratio of the cyclotron frequency and the collisional
frequency. With these assumptions we 
recover the single fluid momentum equation as it is implemented in 
the Bifrost code \citep[see][for details]{Pandey:2008qy}:

\begin{eqnarray}
\rho \left( \frac{\partial }{\partial t}+ {\bf u} \cdot \nabla \right) {\bf u} = - \nabla P + 
{\bf J} \times {\bf B}  \label{eq:bulkmom2}
\end{eqnarray}

\subsubsection{Induction equation}\label{sec:inducapp}

The Ohmic diffusion, Hall term, and ambipolar diffusion are given by

\begin{eqnarray}
\eta_{ohm} = \frac{1}{\sigma} \label{eq:ohmdiff}\\
\eta_{hall} = \frac{ |B|}{q_e n_e}\\
\eta_{amb} = \frac{(|B| \rho_n/\rho)^2}{\rho_i \nu_{in}} = \frac{(|B|
  \rho_n/\rho)^2}{\rho_n \nu_{ni}}
\end{eqnarray}
 
\noindent The electrical conductivity ($\sigma$) in the absence of a magnetic field is

\begin{eqnarray}
\sigma = \frac{q_e^2 n_e}{m_e \nu_e}
 \end{eqnarray}
 
\noindent where the sums of the collision frequencies are written

\begin{eqnarray}
\nu_e = \nu_{en} +\nu_{ei} \\
\nu_i = \nu_{in} +\nu_{ie}
\end{eqnarray}
 
In order to obtain the induction equation the following assumptions 
are made:
\begin{itemize}
\item First, the electric field 

\begin{eqnarray}
{\bf E}+ {\bf u_i} \times {\bf B} = - \frac{\nabla P_e}{n_e q_e} + 
\frac{{\bf J}}{\sigma}+ \frac{\bf J \times B}{q_e n_e}  -\frac{m_e  \nu_{en}}{q_e} {\bf u_D}
\label{eq:etot}
\end{eqnarray}

\noindent is deduced from the electron momentum equation assuming 
zero electron inertia and is expressed in the ion's rest frame.
\item The plasma obeys 
\begin{eqnarray}
\rho_e \nu_{en} << \rho_i \nu_{in}. \label{eq:neg2} 
\end{eqnarray}
\item The term 
\begin{eqnarray}
\frac{\rho_i \rho_n}{\rho}\left[ \frac{d {\bf u_D}}{dt} - ({\bf u_D} \cdot \nabla) {\bf u_i} - 
({\bf u_i} \cdot \nabla) {\bf u_D}\right]\label{eq:neg3}
\end{eqnarray}
\noindent can be neglected when the dynamical frequency of the plasma
is small 
\begin{eqnarray}
\omega \leq \nu_{ni}\rho/\rho_i \label{eq:neg4} 
\end{eqnarray}
\item Biermann's battery contribution, from the $\nabla P_e/q_e n_e$ term
in Equation~\ref{eq:etot}, is neglected. 
\item The term $D\beta_{\mathrm i}/\beta_{\mathrm e}$ is small and of
  order $\leq10^{-3}$, so it is also 
neglected, where $\beta_j=\omega_{cj}/\nu_j$ is the ratio of the 
cyclotron frequency to the sum of the $j$th particle collision frequency.
\item Finally, terms due to the pressure gradient $\nabla P \times {\bf B}$ 
are negligible compared to the induction term ${\bf u \times B}$ when the dynamical 
frequency is small:
\begin{eqnarray}
\omega \leq \left( \frac{v_a^2}{c_s^2}\right)\frac{\rho^2}{\rho_i \rho_n}\nu_{ni}\label{eq:neg6} 
\end{eqnarray}
 \end{itemize}

Under these assumptions the electric field is defined as

\begin{eqnarray}
{\bf E} = \frac{{\bf J}}{\sigma}+ \frac{\bf J \times B}{q_e n_e}  -D^{2}\frac{{\bf J\times B\times B}}{\rho_{i}\nu_{in}},
\label{eq:etotf}
\end{eqnarray}

\noindent and the magnetic field evolution is governed by the
induction equation, derived from the Maxwell equations, and under the considerations listed
above \citep[see][for details]{Parker:2007lr,Pandey:2008qy}.

\begin{eqnarray}
\frac{\partial {\bf B}}{\partial t} = \nabla \times \left[{\bf u \times B} -  \eta {\bf J}
 - \frac{\eta_{hall}}{ |B|} {\bf J \times B} +
\frac{\eta_{amb}}{ B^2} ({\bf J \times B}) \times {\bf B}\right]\label{eq:faradtot2}
\end{eqnarray}

\noindent The right hand side of the induction equation has the convective, Ohmic, 
Hall, and ambipolar terms, from left to right respectively. Note that for simplicity 
we are referring to the Ohmic and ambipolar terms as diffusion terms, but strictly 
speaking none of them can be cast in the form of a diffusion equation
\citep[][already used ambipolar diffusion terminology]{Parker:1963vn}. The two new terms 
(Hall and ambipolar) are implemented in the Bifrost code in the induction equation 
and in the electric field.

Note that from Equation~\ref{eq:faradtot2}, the Hall and ambipolar terms can be 
considered as advection terms:

\begin{eqnarray}
\frac{\partial {\bf B}}{\partial t} =  \nabla \times \left[{\bf u \times B} - \eta {\bf J}
 - {\bf u_H \times B} +{\bf u_A \times B}\right]\label{eq:faradtot3}
\end{eqnarray}

\noindent where the Hall velocity is ${\bf u_H}=(\eta_{hall}{\bf J})/|B| $ 
and the ambipolar velocity is ${\bf u_A}=(\eta_{amb}{\bf J \times B})/B^2 $.

The generalized Ohm's law is implemented in the code using the same 
scheme as used for the MHD equations, i.e., a 6{\it th} order explicit method 
\citep{Gudiksen:2011qy}. From the expression~\ref{eq:faradtot3} it is clear 
that the Hall term and ambipolar diffusivity give rise to two new constraints on the CFL 
condition which restrict the timestep interval \citep{Courant:1928uq} 
($\Delta t_{H}= \Delta x/{\bf u_{H}}$ and $\Delta t_{A}= \Delta x/{\bf u_{A}}$). 
Both velocities are a function of the current ($\nabla\times\mathbf{B}$), i.e., both 
CFL conditions are quadratic functions in $\Delta x$, and the timestep will 
decrease quadratically with increasing spatial resolution. We note that for the 
simulation with mean magnetic field strength in the photosphere of the order 
of 100~G, the ambipolar and Hall velocities are maximal in the cold
regions in the chromosphere with, respectively, 
$u_{A}\approx 100$~km~s$^{-1}$ and  $u_{H}\approx 1$~km~s$^{-1}$. As result of this, the CFL criteria are 
approximately $\Delta t_{A}\approx 0.3$~s and $\Delta t_{H}\approx 20$~s with
$\Delta x\approx 32$~km, compared with the strictest CFL condition in the 
simulation of $\Delta t\approx 3\,10^{-3}$~s. Therefore, as long as we
do not increase the magnetic field and/or the spatial resolution too
much, we do not need to change to an implicit implementation of our equations.

\subsubsection{The energy equation}

As mentioned above, the energy dissipated by Joule heating is given 
by $Q_{Joule} = {\bf E}\cdot{\bf J}$. In the previous section, the Hall term
and ambipolar diffusivity were shown to lead to changes in the
electric field. These changes need to be taken into account when
computing the energy due to the dissipation of the magnetic field.
Note however, that because the Hall term in the electric field is a
function of ${\bf J}\times{\bf B}$, then $({\bf J}\times{\bf B})\cdot
{\bf J}$ is zero, i.e., the Hall term does not produce any energy 
dissipation at all. The only terms which directly dissipate 
electromagnetic energy by dissipation are by Ohmic and by ambipolar
diffusion. In the Bifrost code the former is negligible compared
to the artificial diffusion needed to stabilize the code at
numerically resolvable scales and is therefore set to zero.

In contrast to the artificial resistivity present in the code, the Hall term and 
ambipolar diffusion are calculated as a function of the electron density and, 
for the latter, of the collision frequency between the different species in the 
solar atmosphere. In order to avoid instabilities 
from rapid heating processes due to the new terms, it is sometimes 
necessary to further limit the time steps (beyond the CFL condition) 
because the timescales of the energy dissipation of the ambipolar 
diffusion are short. 
As a result, the energy distribution in the chromosphere changes 
rapidly and the source and sink terms in the energy equation, such as
radiative processes, need to be updated more often than is the case
without ambipolar diffusion.

\subsubsection{Collision frequencies}\label{sec:collfreq}

The collision frequency between electrons and ions can be found in
{\it e.g.} \citet{Priest:1982qy} and is given by

\begin{eqnarray}
\nu_{ei} = 3.759 \, 10^{-6} n_{e} T^{-{3/2}}\ln \Lambda \label{eq:nuei}
\end{eqnarray}

\noindent and

\begin{eqnarray}
\frac{\nu_{en}}{\nu_{ei}} = 5.2 \, 10^{-11} \frac{n_{n}}{n_{e}} \frac{T^{2}}{\ln \Lambda}
\end{eqnarray}

\noindent where $\ln \Lambda$ is the Coulomb logarithm (all in SI units).

As in \citet{De-Pontieu:2001fj}, we follow three different
approximations in computing the collision frequency between 
ions and neutral particles:
as described by \citet{Osterbrock:1961fk,de-Pontieu:1998lr} (hereafter case A), 
as described by \citet{von-Steiger:1989uq} (hereafter case B) and 
as described by \citet{Fontenla:1993fj} (hereafter case C), 
(see Appendix~\ref{app:collfreq}). Table~\ref{tab:runs} lists the 2D simulations 
for which we investigate the effects of these different methods
to calculate $\nu_{in}$. 
We note that the appendix of \citep{De-Pontieu:2001fj} contains two
typos: their formula A6 should be divided by 2 to provide the correct
expression for the collision frequency between neutral hydrogen and
protons, and formula A12 should be replaced by our
formula \label{eq:nuhem2}. 
Our formula \label{eq:nuhp1} provides the correct equation
for the collision frequency between neutral hydrogen and protons,
according to the recipe derived by \citet{de-Pontieu:1998lr} and \citet{Osterbrock:1961fk}.

Throughout the paper we will focus on the results of case B since it
is more recent, and the most extensive.
 
In order to calculate the various collision frequencies, the ion and neutral fractions
are calculated from the Saha-Boltzman equation. The electron density is also 
computed on the basis of LTE, in practice this is done via a table lookup in the 
Bifrost code. In the pre-computed table, the 16 most important atomic species 
in the solar atmosphere are taken into account. Table~\ref{tab:ions} lists the atomic 
species, abundances and ionization fraction ($X_i$).

\subsection{Tests}\label{sec:test}

One of the main objectives of this work is to study the importance and validity of 
the generalized Ohm's law in a ``realistic'' 2.5D simulation of the solar atmosphere.
We describe three different tests done for the implementation of the
generalized Ohm's law, which also illustrate the role and importance of each form of 
diffusivity. For two of the tests, we imposed a velocity equal to zero
at all 
times in the full domain. We also run separate tests using only the Hall term 
or ambipolar diffusion. 

\subsubsection{1D Hall test}

First, we test that our code correctly includes the Hall term. In this test case,
we set the velocities and ambipolar diffusion to zero and consider the 
induction equation in 1D

\begin{eqnarray}
\frac{\partial B_y}{\partial t} = - \eta_{hall} B_x \frac{\partial^2 B_z }{\partial x ^2}\\
\frac{\partial B_z}{\partial t} = \eta_{hall} B_x \frac{\partial^2 B_y }{\partial x ^2}
\end{eqnarray}

\begin{figure}
  \includegraphics[width=0.48\textwidth]{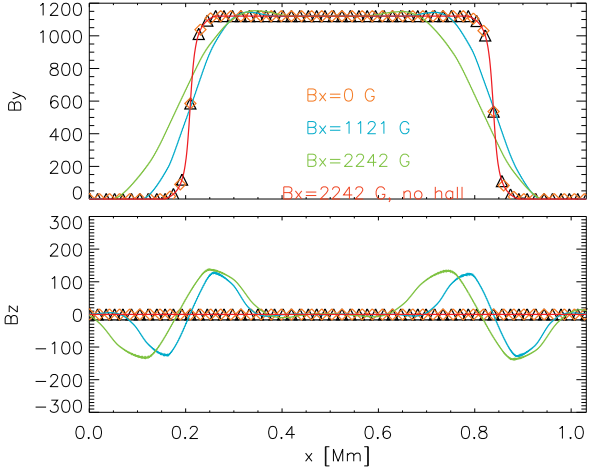}
 \caption{\label{fig:1dhall} $B_y$ (top panel) and $B_z$ (bottom panel) as a function of $x$ 
	are shown for the different 1D simulations with constant Hall term at time $t=20$~s. 
	The initial condition is the same for all simulations (shown with black triangles). The runs have 
	different constant $B_x$ values: $B_x=0$~G (orange diamonds),
        $B_x=1121$~G (blue line), $B_x=2242$~G with the Hall term 
        (green line), and $B_x=2242$~G without the 
	Hall term (red line). Note that the orange diamonds, red line, and black triangles overlap.}
\end{figure}

For this test, we set $B_x$ constant ($B_x= 0,1121,2242$~G are shown with 
orange diamond symbols, and blue and green lines in Figure~\ref{fig:1dhall}). 
With higher $B_{x}$, the rate 
at which $B_y$ and $B_z$ change with time increases. However, the total 
magnetic flux should remain the same at all times, since the Hall term cannot 
convert the magnetic flux into thermal or kinetic energy. Note also that the 
Hall term will give rise to a non-zero $B_z$ (and therefore a non-zero $u_z$ 
in a dynamic simulation) even if the field originally has no component in the
$z$-direction. Figure~\ref{fig:1dhall} shows $B_y$ in the top panel and $B_z$ 
at the bottom panel for four different runs. All cases have the same jump in $B_y$ 
(black triangle symbols in the top panel) and a constant  Hall term. In the test shown in red line
in Figure~\ref{fig:1dhall} does not have the Hall term and $B_{x}=2242$~G.
 All of these tests are shown at the same instant ($t=20$~s). 

The rate of change of  $B_{y}$ and $B_{z}$ is as expected, i.e., the case with 
$B_{x}=2242$~G leads to an increase of unsigned total flux of $B_{z}$
(integrated along the x-axis) that is twice as large as the case where 
$B_{x}=1121$~G. Moreover, the case $B_{x}=0$~G behaves similarly to the 
case with no Hall term. The magnetic flux is in all cases conserved. This gives 
us confidence that our implementation of the Hall term in the code is satisfactory.

\subsubsection{1D Ambipolar test}

In 1D, the induction equation for $B_{y}$ is:

\begin{eqnarray}
\frac{\partial B_y}{\partial t}= \eta_{amb} \frac{\partial}{\partial x}\left( B_y^2 \frac{\partial B_y }{\partial x}\right) \label{eq:1damb}
\end{eqnarray}

\begin{figure}
  \includegraphics[width=0.48\textwidth]{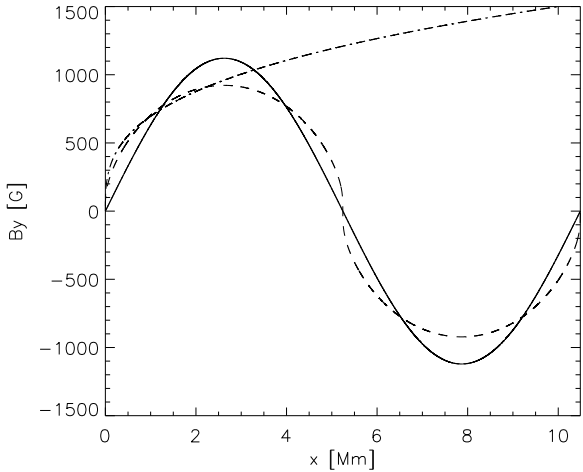}
 \caption{\label{fig:1damb} From the ambipolar test, $B_y$ is shown as a function of 
 	$x$ at t=50~s. The initial condition is shown in 	solid line. The dashed line 
	shows $B_{y}$ at $t=50$~s. The dash dotted line shows a function
	proportional to $x^{1/3}$ which is what would be expected from the analytical 
	considerations.}
\end{figure}

Apart from the trivial solution, $B_y=constant$, it is clear that Equation~\ref{eq:1damb} 
also permits a steady solution of the form of $B_y\propto x^{1/3}$ 
\citep[see][for details]{Brandenburg:1994qy}. In this expression we should keep 
in mind that the code includes numerical diffusivity in addition to ambipolar diffusion. 
We consider the evolution of an initially sinusoidal profile of $B_y$. This profile 
evolves, and strong gradients become stronger approaching the form 
$B_y\propto x^{1/3}$ as time progresses. Figure~\ref{fig:1damb} shows the 
initial condition of $B_y$ (solid line) and at $t=50$~s (dashed line) which is close 
to the steady solution. Observe that where the gradient of $B_y$ is large
$B_{y}$ closely follows the expression $B_y\propto x^{1/3}$ (dash-dotted line). 

Ambipolar diffusion converts magnetic energy into thermal energy as discussed
above. In this 1D test, we turn off the heating from artificial diffusion and only 
allow heating from the ambipolar diffusivity. Such heating in this simple
simulation must follow the expression

\begin{eqnarray}
\frac{\partial e}{\partial t} = \eta_{amb} J_{z}^{2}B_{y}^{2}.
\end{eqnarray}

\begin{figure}
  \includegraphics[width=0.48\textwidth]{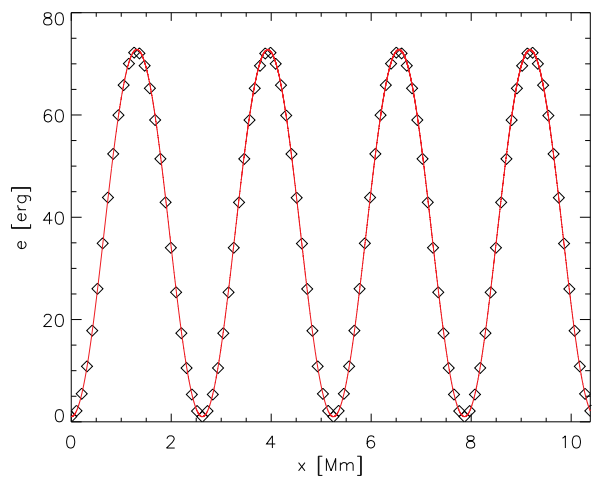}
 \caption{\label{fig:1dambe} Test of ambipolar diffusion on the energy balance. 
 	Energy is shown as a function of $x$ at t=2.1~s. The energy from the model 
	is shown with the black diamonds and the energy extracted from 
	Equation~\ref{eq:enerteo} is shown with the red line. Note that the red line is 
	overlapping with the black diamonds.}
\end{figure}

Figure~\ref{fig:1dambe} shows the energy profile with $x$ of this test (black 
diamonds) at $t=2.1$~s. Calculating the right hand side of this expression using the sinusoidal 
shape of $B_{y}$ from the initial condition, then deriving $J_{z}$, we
calculate the energy to be

\begin{eqnarray}
e= e_{init}+\eta_{amb} J_{z}^{2}B_{y}^{2} \Delta t, \label{eq:enerteo}
\end{eqnarray}

\noindent where $\Delta t$ is the time increment. This relationship is shown
with the red line in Figure~\ref{fig:1dambe}. The black diamonds overlaps the 
red line as would be expected. This indicates we have correctly implemented 
ambipolar diffusion in the code. 

\subsubsection{Collision frequencies test: VAL-C model}

\begin{figure}
  \includegraphics[width=0.48\textwidth]{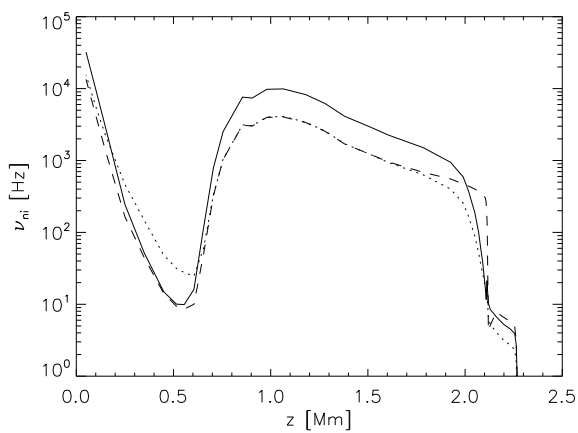}
 \caption{\label{fig:valc} Neutral-ion collision frequency as a function of height for 
 the quiet sun model of \citet{Vernazza:1981yq}, using different formulas for $\nu_{ni}$: the
 dotted line is case A, solid line is case B and dashed line is case C.}
\end{figure}

In order to test whether the absolute values of the diffusion terms are 
calculated correctly, we use three different sources for the 
neutral-ion collision frequency ($\nu_{ni}$, see Section~\ref{sec:collfreq}).  
This also allows us to study the uncertainties involved in the various formulas 
for the collision frequency, as already studied for 1D-static models by \citet{De-Pontieu:2001fj}. 
We test our implementation in the Bifrost code by using the
densities and temperatures from the VAL-C atmospheric model 
\citep{Vernazza:1981yq} which allows us to compare our results with
those found in the published literature. Indeed, we correctly obtain the
neutral-ion collision frequencies as a function of height as can be 
seen by comparing our Figure~\ref{fig:valc} with Figure~2 in
\citet{De-Pontieu:2001fj}. The large dip in the collision frequency at $0.5$~Mm
is due to low number of ions (mostly non-hydrogen species) in this region.. 

\subsection{Initial and boundary conditions}
\label{sec:condition}

\begin{figure}
  \includegraphics[width=0.95\textwidth]{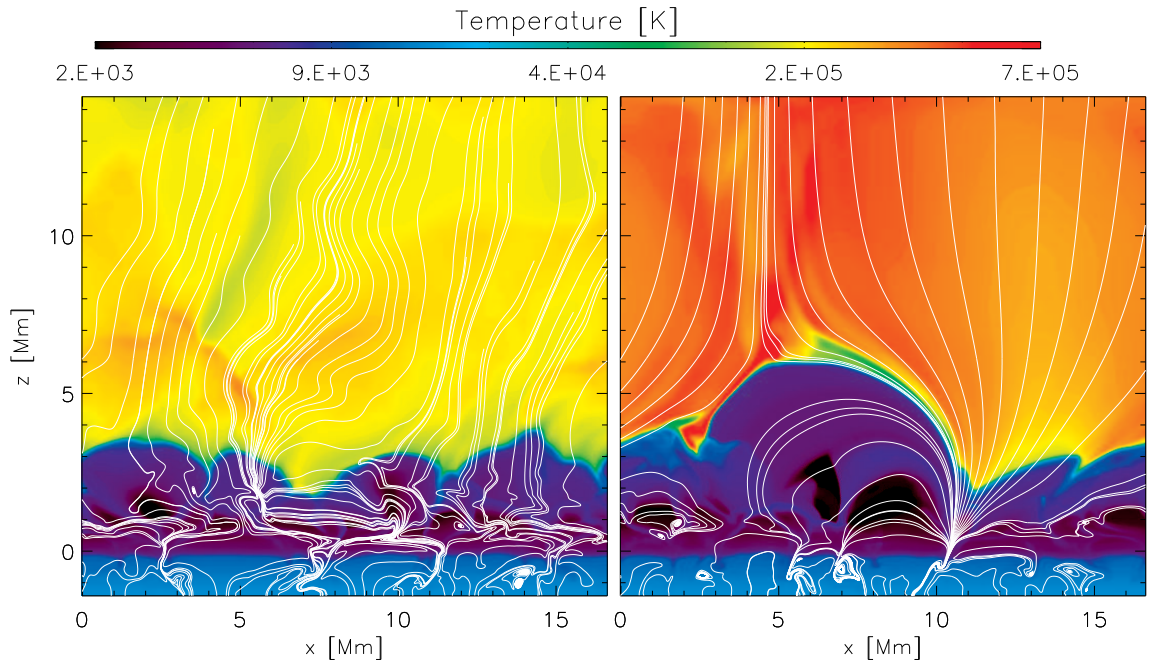}
 \caption{\label{fig:initbox} 2D snapshots of the two initial 2.5D MHD models. 
 	The initial conditions with weak (simulations labeled WA, WB, and WC) 
	and strong magnetic field (SA, SB, and SC) are shown respectively in the 
	left and right panel.
  	The color scale shows the temperature in logarithmic scale and the 
  	magnetic field is shown with white lines.}
\end{figure}

Let us now consider the importance and validity of 
the generalized Ohm's law in a ``realistic'' 2.5D simulation of the solar atmosphere.
The 2.5D computational domain stretches from the upper convection zone to the 
lower corona and is evaluated on a non-uniform grid of $512\times325$
points spanning  $16\times 16$~Mm$^2$.  The frame of reference for the model
is chosen so that $x$ is the horizontal direction and $z$ is the vertical direction 
(Figure~\ref{fig:initbox}). The grid is non-uniform in the 
vertical $z$-direction to ensure that the vertical resolution is good enough to resolve the 
photosphere and the transition region with a grid spacing of $28$~km, while becoming larger at 
coronal heights where gradients are smaller.

We run two different initial conditions with different values for the unsigned magnetic 
field strength but with similar field configurations
(Figure~\ref{fig:initbox}). $B_y$ is originally set to zero.
The initial model starts with a magnetic field that is inclined some
5~degrees with respect to the vertical axis and the two different 
setups for the unsigned field strengths in the photosphere are $0.25$~G, 
the other $90$~G. These two initial conditions are run for the three different 
formulas that were mentioned above to calculate the collision frequency 
$\nu_{in}$. The simulations with the initially weak magnetic field using 
cases A, B or C for the neutral-ion collision frequency are labeled WA, WB 
or WC, while the strong field simulations using cases A, B, or C for
the neutral-ion collision frequency are labeled SA, SB, or SC, respectively.
Table~\ref{tab:runs} list the different simulations.

In the following we will refer in our analysis to simulations WB and SB,
unless otherwise noted.

\section{Results}
\label{sec:results}

The simulations presented include the dynamic processes (including
radiative losses) of the photosphere 
and chromosphere and a self-maintained chromosphere and corona. 
This is a very different type of model as compared to semi-empirical
models such as the VAL-C model, or previous simulations investigating
the effects of partial ionization which had a simplified treatment of
the energy balance (and ionization degree). It is thus of
significant interest to determine how dynamic atmospheres such as 
from our simulations impact the importance of the 
ambipolar diffusion and the Hall terms (also assuming different
approximations to the collision frequency), and to compare the results with
those from the models based on a VAL-C type atmosphere.

The basic structure of our modeled chromospheres are shown in
Figure~\ref{fig:initbox}. A full description of their properties fall
outside the scope of this paper but we will mention the most important
as concerns ambipolar diffusion and the Hall term: the basic thermodynamic state of the
chromosphere is maintained by the continual injection of acoustic
shocks from the photosphere. These perturbations are due to the
chaotic generation of waves in the convection zone, of which waves
with periods of order $3$~minutes will propagate and steepen in the
chromosphere, as is well known and as extensively studied by 
\citet{Carlsson:1992kl,Carlsson+Stein1994}. 
The propagation of waves will be modified in the presence of a
magnetic field \citep[][among others]
{bogdan2003,De-Pontieu:2004hq,Heggland:2007jt,McIntosh:2011fk,Heggland:2011kx}

but will nevertheless steepen and form strong shocks, with high
temperatures in the shock fronts and very low temperatures in the 
regions behind \citep{Leenaarts:2011qy}. These ``cold chromospheric 
bubbles'' can be seen in both panels of Figure~\ref{fig:initbox} 
which show temperatures as low as $2\,000$~K or lower. In the strong 
field case the Lorentz force is clearly important, pushing the corona upwards 
and allowing cool material to exist at great heights, much higher, up to 
5~Mm above the photosphere, than that found in semi-empirical models 
where the maximum chromospheric height is found to be of order 2-2.5~Mm. 
The distribution of density and temperature with height in dynamical 
``realistic''  simulations is discussed in much greater detail in e.g.
\citet{Leenaarts:2011qy}.

\subsection{Collision frequencies and diffusivities}\label{sec:diff}

\begin{figure}
  \includegraphics[width=0.95\textwidth]{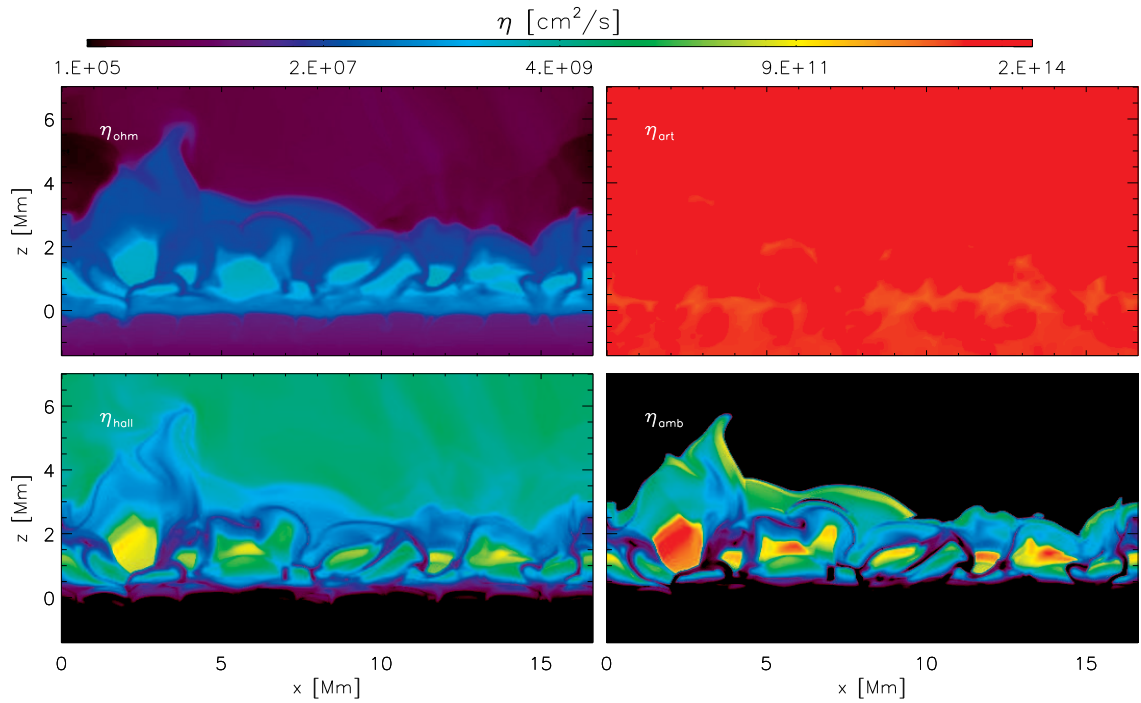}
 \caption{\label{fig:diffL} Comparison of the different  
 diffusivity terms for the simulation WB at $t=500~s$. $\eta_{ohm}$, $\eta_{art}$, 
 $\eta_{hall}$, and $\eta_{amb}$ are shown from top to bottom and left to right 
 respectively in logarithmic scale. Note that more than 9 orders of magnitude are shown.}
\end{figure}

\begin{figure}
  \includegraphics[width=0.95\textwidth]{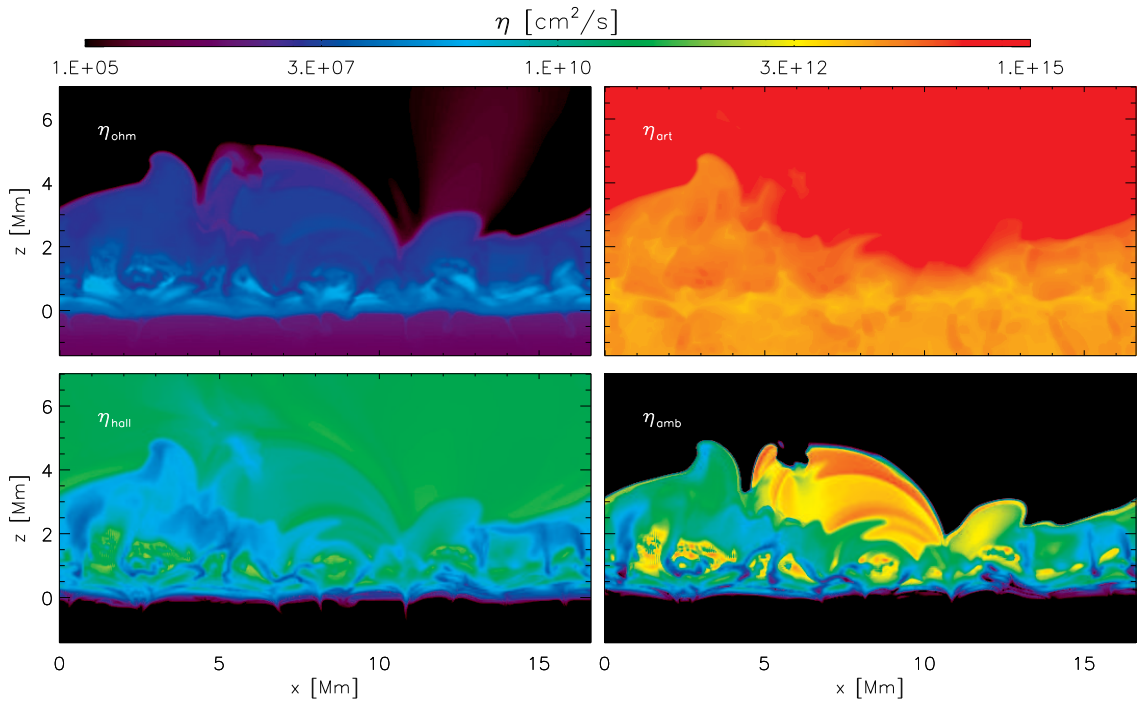}
 \caption{\label{fig:diffS} Comparison of the different 
 diffusivities for the simulation SB at $t=500$~s. The layout is the same as in 
 Figure~\ref{fig:diffL}.}
\end{figure}

As mentioned, most studies of the effects of ion-neutral collisions in the chromosphere 
have been based, in some form, on semi-empirical models (VAL-C or FAL-C models as 
shown in Figure~\ref{fig:valc}). However, the chromosphere and transition 
region are clearly highly dynamic, and it is of great importance to know the effects 
of the neutral-ion interactions in such dynamic atmospheres. 
First of all, we are interested in studying the relative importance of
the different diffusivities in the chromosphere and transition
region. Figures~\ref{fig:diffL} and~\ref{fig:diffS} show the Ohmic diffusion, 
artificial diffusion, Hall term, and ambipolar diffusion from top to bottom and left to right for 
the simulations WB and SB, respectively.

\begin{figure}
  \includegraphics[width=0.95\textwidth]{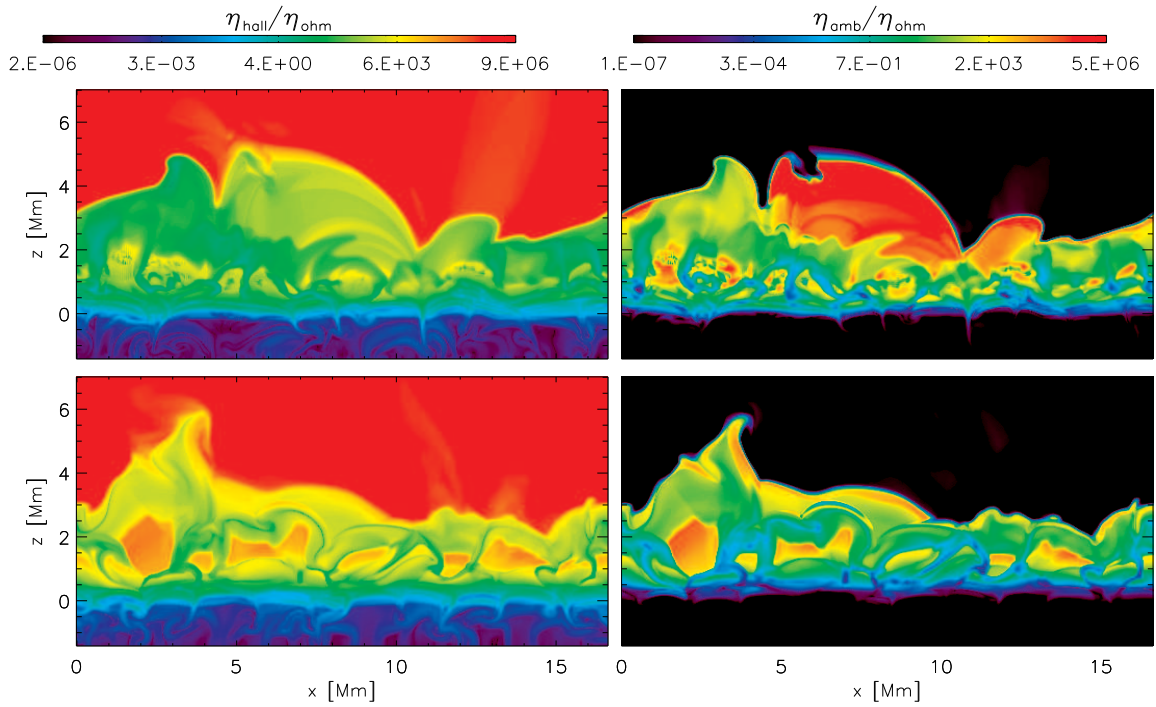}
 \caption{\label{fig:ratio} Ratio between the Hall term and Ohmic diffusion (left panels) and 
 ambipolar and Ohmic diffusion (right panels) for the SB simulation 
 (top panels) and WB (bottom panels) at $t=500$~s.}
\end{figure}

On comparing the different diffusivities, we find that in the entire 
chromosphere, the Hall term is on average two orders of magnitude 
larger than Ohmic diffusion. This is true for both simulations WB and 
SB and is perhaps more easily seen by considering the ratios of the 
diffusivities plotted in Figure~\ref{fig:ratio}. Ambipolar diffusion is 
roughly four orders of magnitude larger in the weak field (WB) case and 
fully six orders of magnitude larger than Ohmic diffusion for the 
strong field SB case. Ambipolar diffusion is considerably larger for 
SB than for WB because ambipolar diffusion depends quadratically on the 
magnetic field strength. Note that while Ohmic diffusion has a 
significant magnitude throughout the 
atmosphere, ambipolar diffusion is important only in the chromosphere. 

Numerical simulations must include some form of artificial diffusion  
in order to compensate for the fact that they do not have infinite 
spatial resolution. In the Bifrost code this is done through a so 
called hyper-diffusivity which in practice means that the diffusion 
coefficient is increased in regions that require high diffusivity, i.e., 
where gradients are large. The magnitude of this artificial diffusion 
is set by the spatial resolution. To some degree this behavior is 
similar to Ohmic diffusion, but there are also significant differences. 
Simulations run at the highest possible spatial resolution cannot even 
come close to the diffusion values found in Ohmic diffusion. This is a
well known problem for numerical simulations of the solar atmosphere.

For the simulations reported here, we find an artificial diffusivity in the 
chromosphere that is three orders of magnitude larger than the Ohmic 
diffusivity and that is 
up to five orders of magnitude larger than the Ohmic diffusivity in the 
corona. By design, the artificial diffusion is largest where the shocks and 
other high gradient phenomena are located. Moreover, since the grid is 
non-uniform and the grid resolution decreases with height, artificial 
diffusion will on average be larger in the corona than in the lower layers. 
In contrast, the Ohmic diffusion is largest in the upper photosphere and 
chromosphere. As result of these differences between artificial diffusion 
and Ohmic diffusion, the magnetic Reynolds number 
on the Sun and in the simulations is completely 
different: the magnetic Reynolds number is several orders of magnitude 
higher in the solar atmosphere than even the highest resolution simulations. 
Since Ohmic diffusion is negligible compared to artificial diffusion we do not 
include its effects either in the induction equation nor in the energy equation. 
In a similar manner, as result of the low resolution of these simulations, 
the artificial diffusion will mask the Hall term, but here we are interested in describing
the stratification of the Hall term and ambipolar diffusion in self-consistent
magneto-convection simulations.

One of the most interesting results of our calculations is that, on the other 
hand, ambipolar diffusion is of the same order or, in some regions, even 
larger than the artificial diffusion in the chromosphere. This is a perhaps 
surprising, but crucial property of the chromosphere. It allows us to use our 
numerical simulations to study the effects of ambipolar diffusion while using 
the correct physical magnitudes of the coefficients. As a result, the 
chromosphere may be the only region where simulations are close to reality 
once all the physics are included in the code, despite the necessarily 
limited resources of todays computing technology. This has an impact 
beyond the chromosphere, since it directly affects discussions on whether 
these self-consistent magneto-convective simulations provide a realistic 
driver and boundary to the corona. For example, recent simulations by 
\citet{Hansteen:2010uq} suggest a preponderance of heating in the lower 
atmosphere (first few Mm above the photosphere), implying that much of 
the coronal heating occurs towards the footpoints 
\citep{Martinez-Sykora:2011fj}. The large ambipolar dissipation we find 
here suggests that such simulations (which only include artificial resistivity) 
are actually much more realistic than previously thought, including the 
predictions of heating low down.

The Ohmic diffusivity, Hall term, and ambipolar diffusivity depend on the electron density, 
while the Ohmic and ambipolar diffusivites also depend on collision 
frequencies  which are shown in Figures~\ref{fig:collL} and~\ref{fig:collS} for 
the weak field WB and strong field SB simulations (see Appendix~\ref{app:collfreq}). 
(Note that the Ohmic diffusivity is proportional to the collision frequency, 
while ambipolar diffusivity is inversely proportional to the collision frequency.) 
On average, the collision frequency between electrons and ions is 
larger than both the ion-neutral, electron-neutral, and neutral-ion collision 
frequencies in the chromosphere, in the WB simulation one order of
magnitude larger and in the SB simulation two orders of magnitude.
This difference in collision frequencies between the WB and SB simulations 
is mainly because the chromosphere is hotter in the SB simulation as a 
result of ambipolar heating. 

The electron-ion collision frequency also shows strong variation
throughout the chromosphere, by almost 
5 orders of magnitude in both simulations. This variation 
is due to the electron density variation in the chromosphere (see second row 
in Figure~\ref{fig:bbr} and Equation~\ref{eq:nuei}). As a result, the electron-ion 
collision frequency is lower inside the cold chromospheric bubbles than in the 
shock fronts. In the chromosphere, the electron-neutral and 
ion-neutral collision frequencies are similar in magnitude. However, 
the ion-neutral collision frequency shows a stronger 
variation in space in the middle chromosphere than the electron-neutral 
collision frequency. This is especially true in the cold chromospheric 
bubbles, where the ion-neutral collision frequency is significantly lower. 
What causes these differences? First, we note that $\rho_{n}$ shows 
less variation in horizontal cuts in the lower chromosphere than
$\rho_{i}$ because the region is mostly dominated by neutrals. 
As result of this, the neutral density is almost similar to the total density.
In the cold bubbles, hydrogen is mostly neutral, and the only ions
are provided by the heavier metals. While both the electron-neutral
and ion-neutral collision frequency are dependent on the neutral
density (which does not vary much in the lower chromosphere), the
dominance of metals in providing ions implies that the average mass
per ion increases significantly in the cold bubbles (compared to
the rest of the chromosphere). The associated drop of average thermal
speed (for the heavy ions compared to protons) is the reason for the
sharp drop in ion-neutral collision frequency in the bubbles (compared
to the rest of the chromosphere). The neutral-ion collision frequency is 
even lower than the ion-neutral collision frequency in the bubbles,
because there are so few ions available to collide with (bottom panels in 
Figure~\ref{fig:bbr}). 

We now consider which parameters are responsible for the changes in diffusivities throughout the 
solar atmosphere. In both simulations (WB and SB), the strongest Ohmic 
diffusivity is concentrated in the lower-middle chromosphere while it is weaker 
in the corona and convection zone. In the chromosphere, the Ohmic diffusivity
varies over a range of almost four orders of magnitude. This variation in 
the chromosphere is due to the strong variation of the electron
density and collision frequency of electrons with neutrals and ions (Figures~\ref{fig:collL}-\ref{fig:bbr}). 
Ohmic diffusion is large in the expanding cool bubbles and low where temperatures are higher. 
This is because the Ohmic diffusion variations are dominated by the
variations in electron density, which is very low in the cool bubbles,
and large in shock fronts. The collision frequency of electrons with ions and neutrals does not
drop as precipitously in the cold bubbles since there are plenty of
neutrals to collide with in these bubbles.

The Hall term is largest in the lower-middle chromosphere and in the corona 
(Figure~\ref{fig:ratio}). This is because it is inversely proportional to the electron 
density which is small in both regions. We see that for both simulations, the Hall 
term is larger than the Ohmic diffusivity in the chromosphere and corona, 
but not in the photosphere nor in the convection zone. In the cooler regions of 
the chromosphere, the Hall term is relatively even higher than in the shock 
fronts, and up to three orders of magnitude greater 
than the Ohmic diffusivity. Such differences are a bit larger in the WB simulation, 
since electron density is smaller in the cold chromospheric 
bubbles in the weak field model. This difference in the electron 
density between WB and SB is because the cold bubbles have cooler 
temperatures in WB simulation than in the simulation SB. 
In the intergranular lanes in the 
photosphere, the Hall term is the most important diffusion term after the 
artificial diffusion. Therefore, since the Hall term is proportional to the
strength of the magnetic field, this term may be important to consider 
in magnetoconvective simulations that include strong magnetic fields. 

Ambipolar diffusion is important in the region from the upper-photosphere 
to the upper-chromosphere. In the photosphere, ambipolar diffusion shows 
some importance in intergranular lanes which have strong concentrations of 
magnetic field (Figure~\ref{fig:bbr}). Therefore, the strong field in the SB 
simulation shows considerably more diffusivity in  intergranules with high 
magnetic flux concentrations than in the weak field WB simulation. In the 
chromosphere, ambipolar diffusion dominates almost everywhere except 
for in the lower chromosphere in shock fronts. The largest difference between 
ambipolar and Ohmic diffusion is located in the cold chromospheric 
bubbles and near the upper-chromosphere/lower transition region. Note 
that for the SB simulation the ratio between ambipolar and Ohmic
diffusion is almost four orders of magnitude larger 
than in the WB simulation due to the quadratic dependence of the ambipolar 
diffusivity on the magnetic field strength. The ambipolar diffusivity is large in 
the cold bubbles since the ion density and the ion-neutral collision frequencies 
are low, but mainly because the ion density is extremely low (5 orders of magnitude 
lower than in the chromospheric shock fronts). In the upper chromosphere, the 
ambipolar diffusivity becomes relatively strong due to low densities --- which 
lead to low ion-neutral collision frequencies --- but only in those regions where 
the magnetic field strength is high. In the cold bubbles the ion density is low 
because of the adiabatic expansion and cooling, whereas in the upper 
chromosphere, it is because the density drops by 2--3 orders
of magnitude compared to the lower chromosphere. 
As shown in Figure~\ref{fig:bbr}, the ambipolar 
diffusivity depends strongly on the ion and neutral density and thus on the 
ionization state of the chromospheric plasma. However, it is well known that in 
the middle and upper chromosphere the ionization and recombination
rates are fairly slow for hydrogen which will not be in ionizational equilibrium 
\citep{Carlsson:2002wl}. This suggests that, in order to treat ambipolar diffusion 
realistically, it is necessary to solve the full time dependent rate equations for 
hydrogen ionization \citep{Leenaarts:2007sf}.

\subsubsection{Comparison with VAL-C model}

The VAL-C model does not provide a good description of the strong temporal and spatial variations 
found in the physical variables of the chromosphere. In the section above, we 
have seen that the different collision frequencies and diffusivities show spatial 
variations of several orders of magnitude at the same height in the chromosphere. 
The lower-chromosphere changes rapidly due to the shock fronts; these lead to 
changes in the thermodynamic structure of the lower chromosphere on  
times-scales shorter than a minute. As a result of this, cold chromospheric bubbles 
appear and disappear in minutes. Due to the ambipolar diffusion, plasma is heated 
in the cold bubbles on timescales shorter than those characterizing the shock front. 
As a result of the spatial and temporal variations, the neutral-ion collision frequency 
varies by almost eight orders of magnitude in the chromosphere in the 2D simulation, 
whereas the VAL-C model has a unique value for the collision frequency at every height
(Figure~\ref{fig:2dvsval}).  In the cold chromospheric bubbles, the collision frequency 
drops to considerably lower values than those found in the VAL-C model. This is a result 
of the low ion number density in these areas, which are overestimated in the VAL-C model. 

We use the maximum, minimum and median magnetic field of the 2D models (SB and WB)
as a function of height in order to calculate the range of ambipolar diffusivities in the semi-empirical 
VAL-C model. The ambipolar diffusion has a very wide range of values, 8 orders of magnitude in 
simulation SB and 11 orders of magnitude for simulation WB. These variations are almost 
6 or 8 orders of magnitude larger than in the VAL-C model. The ambipolar diffusivity
in the 2D simulations is much higher in most of the chromosphere compared to what is found 
in the VAL-C model. The reason for the large difference of the neutral-ion collision
frequency and the ambipolar diffusivity in the VAL-C model (compared
to the 2D model) is because the VAL-C model does not capture the thermodynamics 
of the cold chromospheric bubbles where the neutral-ion collision
frequency drops precipitously. 

These large differences in both the neutral-ion collision frequency and 
ambipolar diffusivities, found between the VAL-C model and our simulation 
should lead to a re-examination of previous results related to the generalized 
Ohm's law (see references) using semi-empirical models to define the density 
and temperature structure. We also reiterate the importance of taking into account 
the likely dynamic state of hydrogen ionization \citep{Leenaarts:2007sf}.

\subsubsection{Other methods to calculate collision frequencies}\label{sec:compcase}

We considered three different methods to calculate $\nu_{in}$ (Section~\ref{sec:collfreq}), 
and thus, the ambipolar diffusivity. Do the different methods give similar values of the 
collision frequency and/or diffusion for the different models? We note that the evolution
of the simulations using the different methods to calculate the collision frequency (WA, 
WB and WC, and SA, SB, and SC) diverge within a few minutes. We therefore integrate 
the properties of the models in time in order to study the different values of the collision 
frequency and ambipolar diffusion, and proceed  as follows. In Figure~\ref{fig:ABCw} 
we show the joint probability distribution function (JPDF) of the temperature versus the 
ion-neutral collision frequency (top) and the ambipolar diffusivity (bottom) for the 
simulations labeled WB (left) and WC (right). In  Figure~\ref{fig:ABCs} we 
show the cases SB and SC in a similar manner. We have integrated over 4 minutes. 
Note that the variation of the $y$-axes are logarithmic and cover more than 10 orders of 
magnitude. We mostly focus on cases B and C since those are the most recent,
and include more advanced calculations of the collision frequencies.

The values for $\nu_{in}$ and $\eta_{amb}$ differ in range and mean values for the 
different cases in each simulation. These differences are significant in certain 
temperature ranges. The differences between the different methods to calculate the 
ion-neutral collision frequency are similar for both atmospheres (weak and strong 
magnetic field strength). For instance, at $\log(T)\approx3.7$ ($5000$~K), case B 
shows ion-neutral collision frequencies that are a factor two larger than for case C. At 
temperatures larger than $\log(T)\approx 3.7$, the collision frequency for case B is almost two orders 
of magnitude smaller than in case C. As a result, in certain temperature ranges, 
the largest values of the ambipolar diffusion for case B are almost  
2 orders of magnitude larger than for case C. 
For temperatures lower than  $\log(T)\approx 3.6$ ($4000$~K), the median collision  
frequency as a function of temperature is roughly similar between cases B and C, 
but not the distribution, as can be seen: case B reaches collision frequencies smaller than 
case C.

At temperatures larger than $\log(T)\approx 3.8$ ($6300$~K), the collision 
frequencies for case C are one order of magnitude larger than for case B.
 As a result, the median of the ambipolar diffusion for cases WC and SC is one 
order of magnitude smaller than for WB and SB.

In order to have a better impression where in the atmosphere the ion-neutral collision 
frequency and ambipolar diffusion differ between the different cases, we take the same 
atmospheric model (simulation WB or SB at $t=2500$~s) and calculate from these 
two models the collision frequency and ambipolar diffusivity using 
the different methods (Figures~\ref{fig:ABCdiffw}-\ref{fig:ABCdiffs}). It is interesting and
important to see that at the precise location where the ambipolar diffusion is really high
(in cold chromospheric bubbles and in the upper chromosphere), the different methods 
differ most. In the cold chromospheric bubbles for both atmospheres (weak and 
strong magnetic field), the collision frequency using case B is almost 4 times smaller than 
case A, but similar to case C. As result, the ambipolar diffusivity is more than 2 times smaller 
using the method of case B than for case A. In the upper chromosphere or shock fronts, 
the collision frequency using case B is almost two times larger than case A 
and slightly larger than with case C. As a result, the 
ambipolar diffusivity using case B is more than 3 times smaller than case A, and 
almost 3 times smaller than case C. However, in the proximity of the transition region, 
the collision frequency for case B is a bit smaller than case A, and more than 10 times 
smaller than case C. As result of this,  the ambipolar diffusivity using case B is a bit 
larger than case A, and more than 10 times larger than case C. 

These large differences between each method are due to the different temperature 
dependences (see Appendix~\ref{app:collfreq}). 
As mentioned, these differences lead to rapidly diverging thermodynamic evolution in 
the various models. Thus, it is important to take into account this uncertainty in calculating 
the collision frequencies when the generalized Ohm's law is modeled.

\subsection{Approximations to the generalized Ohm's law}\label{sec:validation}

The generalized Ohm's law is based on several approximations and 
considerations. In this section we describe where these approximations 
fail and the implications of this failure. We employ the atmospheres of the 
WB and SB simulations in this discussion.

\subsubsection{Approximations in the momentum equation}\label{sec:valmom}

Let us establish and validate the different assumptions underlying the generalized 
Ohm's law as implemented in the code, and see if they are fulfilled in the
fully dynamic self-consistent simulations. One of the first consideration is that 
the ion density dominates over the electron density ($\rho_i/\rho >> \rho_e/\rho$). 
Everywhere in the atmosphere, the values of ion and electron densities remain within the range that fulfill 
$\rho_i/\rho >> \rho_e/\rho$ so that electron inertia can be neglected.

In order to neglect the effects of drift momentum in the momentum equation, 
the drift momentum has to be smaller than the fast momentum ($\rho
\sqrt{v_{a}^{2}+c_{s}^{2}}$, see Equation~\ref{eq:neg1} and \citet{Pandey:2008qy}). This 
approximation is fulfilled in most of the atmosphere under both strong
and weak field conditions. The only exception is in the weak field
atmosphere, where some 
low density areas just below the transition 
region show a ratio of order 0.1-1 (see Figure~\ref{fig:udvacs}). This
is because the ion-neutral
collision frequency drops significantly there, so that
the drift between ions and neutrals becomes rather large. As a result,
in these small regions the plasma becomes decoupled from the neutrals,
and it may 
be necessary to add the drift momentum to the momentum equation,
and/or solve the MHD equations using multiple fluids. 
In the weak field atmosphere, a few of the cold, expanding bubbles show
ratios of order 0.1, so that the fast momentum does stay in excess of
the drift momentum. This suggests that the region below the transition
region is the only one of concern for this particular condition.

\subsubsection{Approximations in the induction equation}

To allow the removal of the time dependence of the drift momentum
equation \citep{Pandey:2008qy}, the electron density times the
collision frequency of electrons with neutrals has to be smaller than the
ion density times the collision frequency of ions with neutrals
( $\rho_e \nu_{en} << \rho_i \nu_{in} $). 
This approximation is fulfilled in most of the atmosphere with the 
exception of some areas in the upper photosphere and in the cold 
chromospheric bubbles (Figure~\ref{fig:rcoll}). In the cold bubbles, the 
electron-neutral collision frequency is almost similar to 
the ion-neutral collision frequency. In the upper photosphere, 
and the collision frequency of electrons with neutrals is 
relatively large so $\rho_e \nu_{en} << \rho_i \nu_{in} $ is not 
fulfilled. Therefore, in these regions, the proper way to solve the ambipolar term in the 
induction equation is by calculating the drift velocity using the fully 
time dependent equation of ${\bf u_D}$ \citep{Pandey:2008qy}.

Some of the approximations used in deriving the equations require that the 
dynamical frequency remains smaller than the frequencies shown in Figure~\ref{fig:omegas}. 
The typical timescales on which the simulated atmosphere evolves is of order 10s or longer, i.e., a
dynamic frequency of $\approx 0.5$~Hz or lower: if the frequencies
shown in Figure~\ref{fig:omegas} are higher than $\approx 0.5$~Hz, the
assumptions underlying the generalized Ohm's law are fulfilled.

The first assumption is that the time derivative of the drift velocity
can be neglected. Following Equation~\ref{eq:neg4}, this can be done only if the dynamical frequency is
smaller than the frequency  $(\rho/\rho_{i})\nu_{ni}$ shown in the top panels of
Figure~\ref{fig:omegas}. The latter frequency is very high in the
upper photosphere and the chromosphere, and stay well above the
dynamical frequency of our simulations ($\approx 0.5$ Hz). Only in the
vicinity of the transition region does  $(\rho/\rho_{i})\nu_{ni}$
become small enough that it is of the same order as the dynamical
frequency of the simulations. As a result, we may need to take into
account the derivative terms shown in Equation~\ref{eq:neg3} only in this small region 
in the vicinity of the transition region. 

A second assumption is that the dyadic product of the drift velocity
in the momentum equation can be neglected \citep{Pandey:2008qy}. This
term can only be neglected if the
dynamic frequency stays well below the frequency defined in
Equation~\ref{eq:neg5} and shown in the middle panels in
Figure~\ref{fig:omegas}. We find that in the cold chromospheric
bubbles (in the weak field case) and in the upper chromosphere (in
both weak and strong field cases), this assumption sometimes fails. In
these regions, we may thus need to take into account the momentum drift term in the momentum 
equation. 

A final assumption is that the terms of the form $\nabla P \times {\bf B}$ in the
induction equation can be neglected. This can only be done when the
dynamic frequency stays below the frequency defined in Equation~\ref{eq:neg6} and 
shown in the bottom panels in Figure~\ref{fig:omegas}. This bound for the dynamical 
frequency strongly depends on magnetic field strength of the model. 
We find that for the weakly magnetic atmosphere case (WB) this limit
is low, and the assumption fails in the upper chromosphere and cold
bubbles.In the strongly magnetic atmosphere (SB) the assumptions only fails in
the upper part of the chromosphere.

In summary, for the weak field atmosphere we cannot neglect the time
derivative and dyadic product of the
drift velocity, and the $\nabla P \times {\bf B}$ terms (in the
momentum and induction equation respectively) in the cold bubbles and
just below the transition region. In all other regions in the weak
field atmosphere, the assumptions underlying the generalized Ohm's law are fulfilled.

For the strong field atmosphere, the generalized Ohm's law works well
in most of the chromosphere, except in the region just below the
transition region where the time derivative and dyadic product of the
drift velocity and  the $\nabla P \times {\bf B}$ terms cannot be neglected.

\clearpage

\section{Discussion and Conclusions}\label{sec:conclusions}

We have implemented the partial ionization effects in the Bifrost code in the form of the 
Hall term and ambipolar diffusion. The code has been tested and verified with different 
tests that are presented in this paper. The code allows the simulation of the solar 
atmosphere, from the upper convection zone to the lower corona, with a 
magnetoconvective photosphere, and a fully-dynamic and self-maintained chromosphere 
and corona. We studied the different diffusivities in two different models, one is weakly 
magnetic, and the other is rather strongly magnetic. The magnetic field strength 
of the latter model is similar to that found in the quiet sun,
including the network. 

In short, the Ohmic diffusion is roughly three orders of magnitude smaller than the 
Hall term in the chromosphere, and the latter is three orders of magnitude smaller 
than the artificial diffusion. Unlike Ohmic diffusion, the Hall term depends on the 
magnetic field, as does ambipolar diffusion which is strongly dependent on the 
magnetic field strength. As a result of this, the ambipolar diffusivity is clearly different 
for the two models; in regions with large ambipolar diffusivity we find it is of the same 
order as the artificial diffusion in the chromosphere for the weakly magnetic 
model (WB), and more than one order of magnitude larger than the artificial diffusivity 
for the strongly magnetic model (SB). The fact that the artificial
diffusivity is actually smaller than the ambipolar diffusivity under many
chromospheric conditions has some very important consequences. It
means that these simulations are
capable of providing a surprisingly realistic view of 
the consequences of the ambipolar diffusion in the chromosphere and 
corona. This has an impact beyond the chromosphere, since it directly affects
discussions on whether these self-consistent magneto-convective
simulations provide a realistic driver and boundary to the corona. These 
results will be described in detail in a follow up paper. 

Another important result is that both, the Hall term and ambipolar diffusivity, 
vary by several orders of magnitude in the chromosphere 
as result of the time varying dynamics and the strong variations in temperature, electron, 
ion and neutral density, and magnetic field strength in this region. This strong 
variation is not taken into account in any of the previous studies
which use either 1D 
semi-empirical VAL-C type models, or lack more sophisticated
approaches to the radiation, ionization and energy balance. 
The largest values of the ambipolar diffusivity are located
in the cold chromospheric bubbles that have low temperatures due to strong adiabatic 
expansion, and in the upper chromosphere because the neutral-ion collision 
frequency is small. However, the ambipolar diffusion is strongly dependent on the ionization 
degree, and as shown by \citet{Leenaarts:2007sf}, time dependent hydrogen will change 
the ratio between neutrals and ions compared to LTE conditions. The Bifrost code can treat the 
time-dependent ionization of hydrogen and we plan to run new simulations
taking into account both the generalized Ohm's law and time-dependent 
hydrogen ionization. 

We have compared different methods to calculate the collision frequency between 
neutrals and ions. Both the ion-neutral collision frequency and ambipolar diffusivity 
differ considerably as a function of the method used to calculate this collision 
frequency. Since ambipolar diffusion has a significant impact on the
thermodynamic evolution of these models, the 
simulations rapidly diverge. When comparing each method we find the largest 
differences are located in regions where the ambipolar diffusivity is large: in the cold 
chromospheric bubbles and in the upper chromosphere in the vicinity of the transition 
region. These differences bring a new uncertainty to the results
(Section~\ref{sec:compcase}), and highlight the need for a detailed
consideration of the relevant collisional processes in the chromosphere.

Finally, we investigated the different approximations underlying the
generalized Ohm's law as described in detail by \citet{Pandey:2008qy}. In both models, 
most of the simplifications are applicable with some exceptions. In the upper-chromosphere 
the collision frequency is too low, as a consequence, the velocity drift can be large. 
Therefore, we may need to define the velocity drift and add an extra term in the momentum 
equation related to the momentum drift between ions and neutrals. In the upper photosphere, 
and in cold chromospheric bubbles the ambipolar term in the induction equation may need 
to be calculated using the drift velocity. Moreover, the drift velocity should be calculated using 
the time dependent form \citep[as shown in][]{Pandey:2008qy}. This is necessary because the 
ion density and the ion-neutrals collision frequency drop in these cold areas as opposed to 
the electron density and the electron-neutral collision frequency. 

\section{Acknowledgments}

The 2D simulations have been run with the Njord and Stallo cluster from the Notur
project, and the Pleiades cluster through computing grants SMD-07-0434, SMD-08-0743, 
SMD-09-1128, SMD-09-1336, SMD-10-1622, SMD-10-1869, SMD- 11-2312, and 
SMD-11-2752 from the High End Computing (HEC) division of NASA. 
We thankfully acknowledge the computer and supercomputer 
resources of the Research Council of Norway through grant 170935/V30 and through 
grants of computing time from the Programme for Supercomputing. 
This work has benefited from discussions at the International Space Science Institute 
(ISSI) meeting on ``Heating of the magnetized chromosphere'' from 21-24 February, 2012, 
where many aspects of this paper were discussed with other colleagues.
To analyze the data we have used IDL. B.D.P. was supported through NASA grants 
NNX08BA99G, NNX08AH45G and NNX11AN98G.

\bibliographystyle{aa}

\begin{thebibliography}{}

\bibitem[\protect\citeauthoryear{{Arber}, {Haynes}, \& {Leake}}{{Arber} et
  al.}{2007}]{Arber:2007yf}
{Arber} T.~D., {Haynes} M.,  {Leake} J.~E., 2007, \apj, 666, 541

\bibitem[\protect\citeauthoryear{{Bogdan} et al.}{{Bogdan} et
  al.}{2003}]{bogdan2003}
{Bogdan} T.~J., {Carlsson} M., {Hansteen} V.~H., et al., 2003, \apj, 599, 626

\bibitem[\protect\citeauthoryear{{Brandenburg} \& {Zweibel}}{{Brandenburg} \&
  {Zweibel}}{1994}]{Brandenburg:1994qy}
{Brandenburg} A.,  {Zweibel} E.~G., 1994, \apjl, 427, L91

\bibitem[\protect\citeauthoryear{{Carlsson} \& {Leenaarts}}{{Carlsson} \&
  {Leenaarts}}{2012}]{Carlsson:2012uq}
{Carlsson} M.,  {Leenaarts} J., 2012, ArXiv e-prints, in press in ApJ

\bibitem[\protect\citeauthoryear{{Carlsson} \& {Stein}}{{Carlsson} \&
  {Stein}}{1992}]{Carlsson:1992kl}
{Carlsson} M.,  {Stein} R.~F., 1992, \apjl, 397, L59

\bibitem[\protect\citeauthoryear{{Carlsson} \& {Stein}}{{Carlsson} \&
  {Stein}}{1994}]{Carlsson+Stein1994}
{Carlsson} M.,  {Stein} R.~F., 1994, {Radiation shock dynamics in the solar
  chromosphere - results of numerical simulations}, en {Carlsson} M. (ed.),
  Chromospheric Dynamics, p.~47

\bibitem[\protect\citeauthoryear{{Carlsson} \& {Stein}}{{Carlsson} \&
  {Stein}}{1997}]{Carlsson:1997tg}
{Carlsson} M.,  {Stein} R.~F., 1997, \apj, 481, 500

\bibitem[\protect\citeauthoryear{{Carlsson} \& {Stein}}{{Carlsson} \&
  {Stein}}{2002}]{Carlsson:2002wl}
{Carlsson} M.,  {Stein} R.~F., 2002, \apj, 572, 626

\bibitem[\protect\citeauthoryear{{Cheung} \& {Cameron}}{{Cheung} \&
  {Cameron}}{2012}]{Cheung:2012vn}
{Cheung} M.~C.~M.,  {Cameron} R.~H., 2012, ArXiv e-prints, in press


\bibitem[\protect\citeauthoryear{{Courant}, {Friedrichs}, \& {Lewy}}{{Courant}
  et al.}{1928}]{Courant:1928uq}
{Courant} R., {Friedrichs} K.,  {Lewy} H., 1928, Mathematische Annalen, 100,
  32, {{\"U}ber die partiellen Differenzengleichungen der mathematischen
  Physik}

\bibitem[\protect\citeauthoryear{{Cowling}}{{Cowling}}{1957}]{cowling1957}
{Cowling} T.~G., 1957, Magnetohydrodinamics.
\newblock Interscience tracts on physics and astronomy

\bibitem[\protect\citeauthoryear{{De Pontieu}}{{De
  Pontieu}}{1999}]{de-Pontieu:1999uq}
{De Pontieu} B., 1999, \aap, 347, 696

\bibitem[\protect\citeauthoryear{{De Pontieu}, {Erd{\'e}lyi}, \& {James}}{{De
  Pontieu} et al.}{2004}]{De-Pontieu:2004hq}
{De Pontieu} B., {Erd{\'e}lyi} R.,  {James} S.~P., 2004, \nat, 430, 536

\bibitem[\protect\citeauthoryear{{De Pontieu} \& {Haerendel}}{{De Pontieu} \&
  {Haerendel}}{1998}]{de-Pontieu:1998lr}
{De Pontieu} B.,  {Haerendel} G., 1998, \aap, 338, 729

\bibitem[\protect\citeauthoryear{{De Pontieu}, {Martens}, \& {Hudson}}{{De
  Pontieu} et al.}{2001}]{De-Pontieu:2001fj}
{De Pontieu} B., {Martens} P.~C.~H.,  {Hudson} H.~S., 2001, \apj, 558, 859

\bibitem[\protect\citeauthoryear{{Erd{\'e}lyi} \& {James}}{{Erd{\'e}lyi} \&
  {James}}{2004}]{Erdelyi:2004qy}
{Erd{\'e}lyi} R.,  {James} S.~P., 2004, \aap, 427, 1055

\bibitem[\protect\citeauthoryear{{Fontenla}, {Avrett}, \& {Loeser}}{{Fontenla}
  et al.}{1990}]{Fontenla:1990yq}
{Fontenla} J.~M., {Avrett} E.~H.,  {Loeser} R., 1990, \apj, 355, 700

\bibitem[\protect\citeauthoryear{{Fontenla}, {Avrett}, \& {Loeser}}{{Fontenla}
  et al.}{1993}]{Fontenla:1993fj}
{Fontenla} J.~M., {Avrett} E.~H.,  {Loeser} R., 1993, \apj, 406, 319

\bibitem[\protect\citeauthoryear{{Geiss} \& {Buergi}}{{Geiss} \&
  {Buergi}}{1986}]{Geiss:1986fk}
{Geiss} J.,  {Buergi} A., 1986, \aap, 159, 1

\bibitem[\protect\citeauthoryear{{Goodman}}{{Goodman}}{2000}]{Goodman:2000ys}
{Goodman} M.~L., 2000, \apj, 533, 501

\bibitem[\protect\citeauthoryear{{Gudiksen} et al.}{{Gudiksen} et
  al.}{2011}]{Gudiksen:2011qy}
{Gudiksen} B.~V., {Carlsson} M., {Hansteen} V.~H., et al., 2011, \aap, 531,
  A154

\bibitem[\protect\citeauthoryear{{Hansteen} et al.}{{Hansteen} et
  al.}{2010}]{Hansteen:2010uq}
{Hansteen} V.~H., {Hara} H., {De Pontieu} B.,  {Carlsson} M., 2010, \apj, 718,  1070

\bibitem[\protect\citeauthoryear{{Heggland}, {De Pontieu}, \&
  {Hansteen}}{{Heggland} et al.}{2007}]{Heggland:2007jt}
{Heggland} L., {De Pontieu} B.,  {Hansteen} V.~H., 2007, \apj, 666, 1277


\bibitem[\protect\citeauthoryear{{Heggland} et al.}{{Heggland} et
  al.}{2011}]{Heggland:2011kx}
{Heggland} L., {Hansteen} V.~H., {De Pontieu} B.,  {Carlsson} M., 2011, \apj,
  743, 142

\bibitem[\protect\citeauthoryear{{Hyman}, {Vichnevtsky}, \&
  {Stepleman}}{{Hyman} et al.}{1979}]{Hyman1979}
{Hyman} J., {Vichnevtsky} R.,  {Stepleman} R., 1979, Adv. in Comp. Meth,
  PDE's-III, 313

\bibitem[\protect\citeauthoryear{{James} \& {Erd{\'e}lyi}}{{James} \&
  {Erd{\'e}lyi}}{2002}]{James:2002lr}
{James} S.~P.,  {Erd{\'e}lyi} R., 2002, \aap, 393, L11

\bibitem[\protect\citeauthoryear{{James}, {Erd{\'e}lyi}, \& {De
  Pontieu}}{{James} et al.}{2003}]{James:2003fk}
{James} S.~P., {Erd{\'e}lyi} R.,  {De Pontieu} B., 2003, \aap, 406, 715

\bibitem[\protect\citeauthoryear{{Khodachenko} et al.}{{Khodachenko} et
  al.}{2004}]{Khodachenko:2004vn}
{Khodachenko} M.~L., {Arber} T.~D., {Rucker} H.~O.,  {Hanslmeier} A., 2004,
  \aap, 422, 1073

\bibitem[\protect\citeauthoryear{{Khodachenko} et al.}{{Khodachenko} et
  al.}{2006}]{Khodachenko:2006kx}
{Khodachenko} M.~L., {Rucker} H.~O., {Oliver} R., {Arber} T.~D.,  {Hanslmeier}
  A., 2006, Advances in Space Research, 37, 447

\bibitem[\protect\citeauthoryear{{Khomenko} \& {Collados}}{{Khomenko} \&
  {Collados}}{2012}]{Khomenko:2012ys}
{Khomenko} E.,  {Collados} M., 2012, \apj, 747, 87

\bibitem[\protect\citeauthoryear{{Leake} \& {Arber}}{{Leake} \&
  {Arber}}{2006}]{Leake:2006kx}
{Leake} J.~E.,  {Arber} T.~D., 2006, \aap, 450, 805

\bibitem[\protect\citeauthoryear{{Leake}, {Arber}, \& {Khodachenko}}{{Leake} et
  al.}{2005}]{Leake:2005rt}
{Leake} J.~E., {Arber} T.~D.,  {Khodachenko} M.~L., 2005, \aap, 442, 1091

\bibitem[\protect\citeauthoryear{{Leenaarts} et al.}{{Leenaarts} et
  al.}{2011}]{Leenaarts:2011qy}
{Leenaarts} J., {Carlsson} M., {Hansteen} V.,  {Gudiksen} B.~V., 2011, \aap,
  530, A124

\bibitem[\protect\citeauthoryear{{Leenaarts} et al.}{{Leenaarts} et
  al.}{2007}]{Leenaarts:2007sf}
{Leenaarts} J., {Carlsson} M., {Hansteen} V.,  {Rutten} R.~J., 2007, \aap, 473,
  625

\bibitem[\protect\citeauthoryear{{Mart{\'{\i}}nez-Sykora} et
  al.}{{Mart{\'{\i}}nez-Sykora} et al.}{2011}]{Martinez-Sykora:2011fj}
{Mart{\'{\i}}nez-Sykora} J., {De Pontieu} B., {Hansteen} V.,  {McIntosh} S.~W.,
  2011, \apj, 732, 84

\bibitem[\protect\citeauthoryear{{McIntosh} et al.}{{McIntosh} et
  al.}{2011}]{McIntosh:2011fk}
{McIntosh} S.~W., {De Pontieu} B., {Carlsson} M., et al., 2011, \nat, 475, 477

\bibitem[\protect\citeauthoryear{{Nordlund}}{{Nordlund}}{1982}]{Nordlund1982}
{Nordlund} {\AA}., 1982, Aap, 107, 1

\bibitem[\protect\citeauthoryear{{Osterbrock}}{{Osterbrock}}{1961}]{Osterbrock:1961fk} {Osterbrock} D.~E., 1961, \apj, 134, 347

\bibitem[\protect\citeauthoryear{{Pandey} \& {Wardle}}{{Pandey} \&
  {Wardle}}{2008}]{Pandey:2008qy}
{Pandey} B.~P.,  {Wardle} M., 2008, \mnras, 385, 2269

\bibitem[\protect\citeauthoryear{{Parker}}{{Parker}}{1963}]{Parker:1963vn}
{Parker} E.~N., 1963, \apjs, 8, 177

\bibitem[\protect\citeauthoryear{{Parker}}{{Parker}}{2007}]{Parker:2007lr}
{Parker} E.~N., 2007, {Conversations on Electric and Magnetic Fields in the
  Cosmos}.
\newblock Princeton University Press

\bibitem[\protect\citeauthoryear{{Priest}}{{Priest}}{1982}]{Priest:1982qy}
{Priest} E.~R., 1982, {Solar magneto-hydrodynamics}.
\newblock p. 74P

\bibitem[\protect\citeauthoryear{{Schaffenberger} et al.}{{Schaffenberger} et
  al.}{2005}]{Schaffenberger:2005fj}
{Schaffenberger} W., {Wedemeyer-B{\"o}hm} S., {Steiner} O.,  {Freytag} B.,
  2005, {Magnetohydrodynamic Simulation from the Convection Zone to the
  Chromosphere}, en ESA Special Publication, Vol. 596, {D.~E.~Innes, A.~Lagg,
  \& S.~A.~Solanki}  (ed.), Chromospheric and Coronal Magnetic Fields

\bibitem[\protect\citeauthoryear{{Skartlien}}{{Skartlien}}{2000}]{Skartlien2000}{Skartlien} R., 2000, \apj, 536, 465

\bibitem[\protect\citeauthoryear{{Stein} \& {Nordlund}}{{Stein} \&
  {Nordlund}}{2006}]{Stein:2006qy}
{Stein} R.~F.,  {Nordlund} {\AA}., 2006, \apj, 642, 1246 

\bibitem[\protect\citeauthoryear{{Vernazza}, {Avrett}, \& {Loeser}}{{Vernazza}
  et al.}{1981}]{Vernazza:1981yq}
{Vernazza} J.~E., {Avrett} E.~H.,  {Loeser} R., 1981, \apjs, 45, 635

\bibitem[\protect\citeauthoryear{{V{\"o}gler} et al.}{{V{\"o}gler} et
  al.}{2005}]{Vogler:2005fj}
{V{\"o}gler} A., {Shelyag} S., {Sch{\"u}ssler} M., et al., 2005, \aap, 429,  335

\bibitem[\protect\citeauthoryear{{von Steiger} \& {Geiss}}{{von Steiger} \&
  {Geiss}}{1989}]{von-Steiger:1989uq}
{von Steiger} R.,  {Geiss} J., 1989, \aap, 225, 222

\end{thebibliography}

\appendix
\section{Collision frequencies}\label{app:collfreq}

In order to calculate the collision frequency between ions and 
neutral particles we use three different approximations \citep[following the
approach by][]{De-Pontieu:2001fj}: 
one described by \citet{Osterbrock:1961fk} (hereafter case A), 
one described by \citet{von-Steiger:1989uq} 
(hereafter case B) and one by \citet{Fontenla:1993fj} (hereafter case C). 

As a first approach (case A), we
take the formulas from \citet{Osterbrock:1961fk} and \citet{de-Pontieu:1998lr}, 
where the collision frequency between neutral hydrogen and 
protons ($\nu_{Hp}$) are given by:

\begin{eqnarray}
\nu_{Hp} = 5 \, 10^{-19} \sqrt{\frac{1}{2}}\sqrt{\frac{8 k T}{\pi m_H}} n_p \label{eq:nuhp1}
\end{eqnarray}

\noindent $m_H$ is the hydrogen atom mass, and $n_p$ is the proton number density. 
Note that \citet{De-Pontieu:2001fj} had a typo with a factor of 2. 
The collision frequency ($\nu_{Hm}$) of neutral hydrogen with an ionized 
metal is defined as:

\begin{eqnarray}
\nu_{Hm} = 8 \, 10^{-20} \sqrt{\frac{m_m}{m_m+1}}\sqrt{\frac{8 k T}{\pi m_H}} n_m
\end{eqnarray}

\noindent where $m_m$ and $n_m$ are the atomic mass number of metals ions 
and the number density of metals ions of type $m$, respectively. The collisions between 
neutral helium and ions is given by:

\begin{eqnarray}
\nu_{Hep} = 4 \, 10^{-20} \sqrt{\frac{1}{5}}\sqrt{\frac{8 k T}{\pi m_H}} n_p \\
\nu_{Hep} = 4 \, 10^{-20} \sqrt{\frac{m_m}{m_m+1}}\sqrt{\frac{8 k T}{\pi m_H}} n_m
\end{eqnarray}

For the second approach (Case B), following \citet{De-Pontieu:2001fj}, 
\citet{von-Steiger:1989uq} describe the collision rate as follows:

\begin{eqnarray}
\nu_{Hp} = 118 \sqrt{\frac{T}{10^4}}\left(1 - 0.125 \log{\frac{T}{10^4}}\right)^2\frac{n_p}{10^{16}} \\
\nu_{Hm} = 21.05 \sqrt{\frac{A_m}{A_m+1}} Z_m \frac{n_m}{10^{16}} 
\end{eqnarray}

For the helium-proton and helium-metal collision frequency we follow \citet{Geiss:1986fk}:

\begin{eqnarray}
\nu_{Hep} = 2.2 \frac{n_p/10^6}{\sqrt{T/10^4}} Z_m \\
\nu_{Hem} = 5.84\sqrt{\frac{A_m}{A_m+1}}  Z_m \frac{n_m}{10^{16}} \label{eq:nuhem2}
\end{eqnarray}

\noindent where $Z_m$ is the ionization weight and we considered that the ions have only one 
ionization state, i.e., $Z_m=1$.  Note that \citet{De-Pontieu:2001fj} have a typo where the 
expression for $\nu_{Hep}$ is missing the square root symbol for the temperature and the constant 
$2.2$ is also different. 

Finally, we find the collision frequencies for the Case C in the appendix of \citet{Fontenla:1993fj}.

Using these collision frequencies  (Eq.~\ref{eq:nuhp1}-\ref{eq:nuhem2}), the collision frequency 
of neutral Hydrogen with all ions is given by: 

\begin{eqnarray}
\nu_{Hi} &  = & \nu_{Hp} + \nu_{HC} + \nu_{HN} + \nu_{HO} + \nu_{HNe} + \nu_{HNa} + \nu_{HMg} + \nu_{HAl}  \\
&& + \nu_{HSi} + \nu_{HS} + \nu_{HK}  + \nu_{HCa} + \nu_{HCr} + \nu_{HFe} + \nu_{HNi}  
\end{eqnarray}

\noindent and similarly for the collision frequency of neutral Helium with all ions. Finally, the 
average neutral-ion collision frequency is given by 

\begin{eqnarray}
\nu_{ni} = \frac{\rho_H}{\rho_n}\nu_{Hi} + \frac{\rho_{He}}{\rho_n}\nu_{Hei}
\end{eqnarray}

Note that in the main text we often use $\nu_{in}$, which can be derived from $\nu_{ni}$ using 
momentum conservation ($\rho_j \nu_{jk} = \rho_k \nu_{kj}$).

\clearpage

\begin{figure}
  \includegraphics[width=0.95\textwidth]{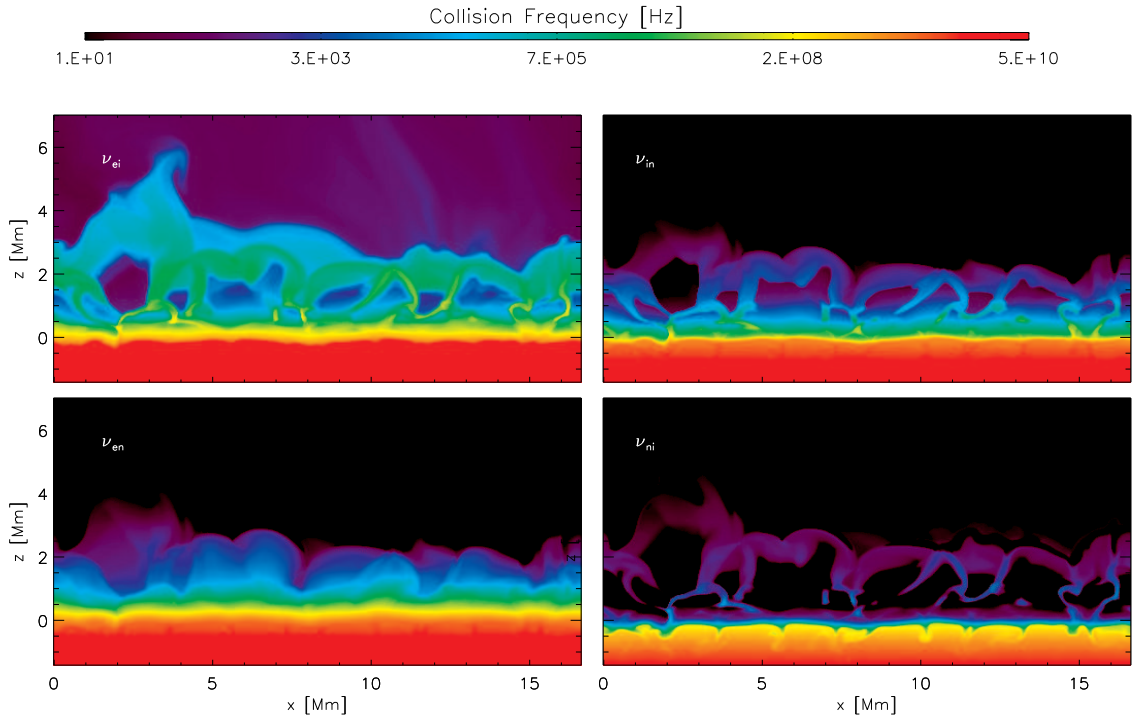}
 \caption{\label{fig:collL} Comparison of the different collision 
 frequencies for the WB simulation at $t=500~s$. $\nu_{ei}$, $\nu_{en}$, $\nu_{in}$, 
 and $\nu_{ni}$ are shown from top to bottom and left to right respectively in logarithmic scale.} 
\end{figure}

\begin{figure}
  \includegraphics[width=0.95\textwidth]{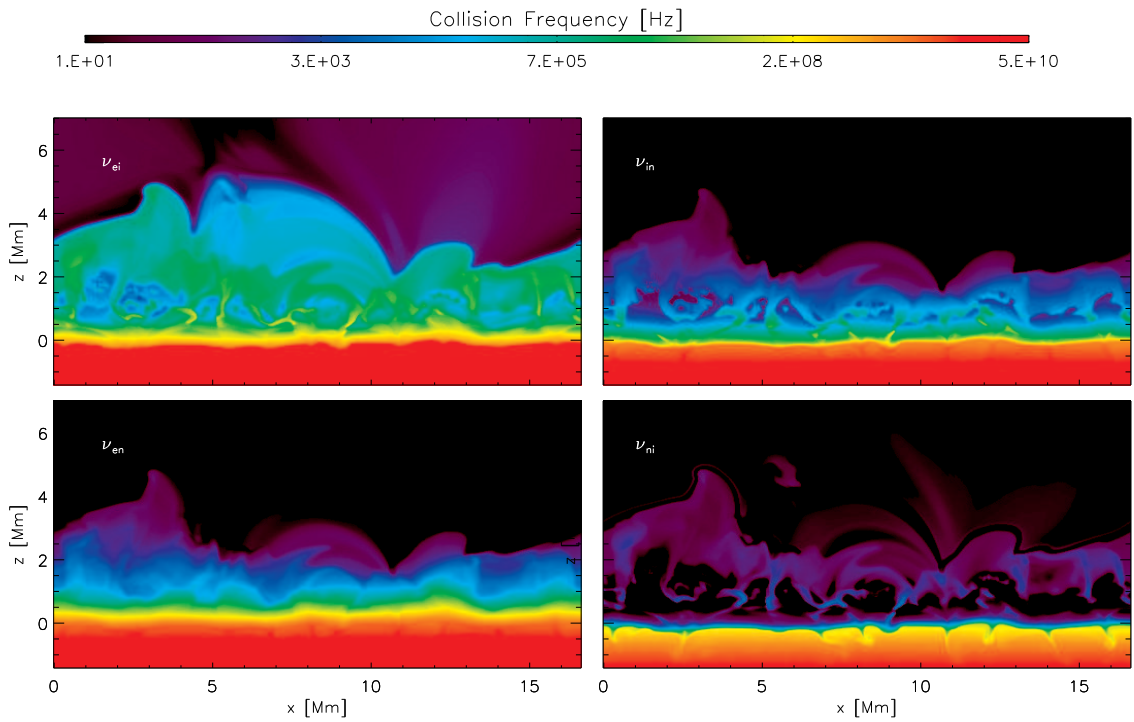}
 \caption{\label{fig:collS} Comparison of the different collision 
 frequencies for the SB simulation at $t=500$~s. The layout is the same as in 
 Figure~\ref{fig:collL}.}
\end{figure}

\begin{figure}
  \includegraphics[width=0.95\textwidth]{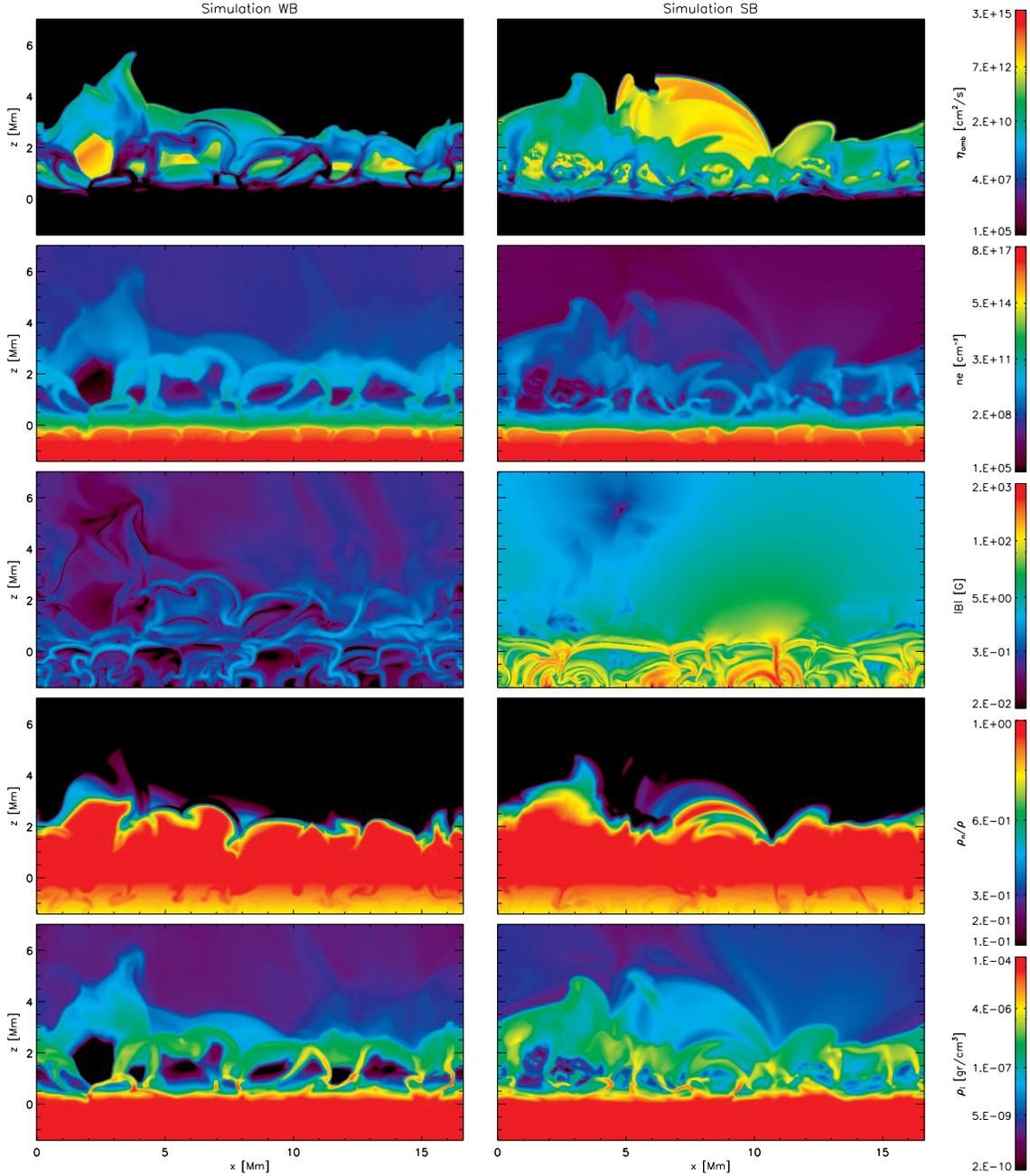}
 \caption{\label{fig:bbr} The ambipolar diffusion, electron density, absolute value of the 
 magnetic field, ratio between neutral and total density and ion density are shown from
 top to bottom for the WB simulation (left panels) and SB (right panels) at $t=500$~s.}
\end{figure}

\begin{figure}
  \includegraphics[width=0.95\textwidth]{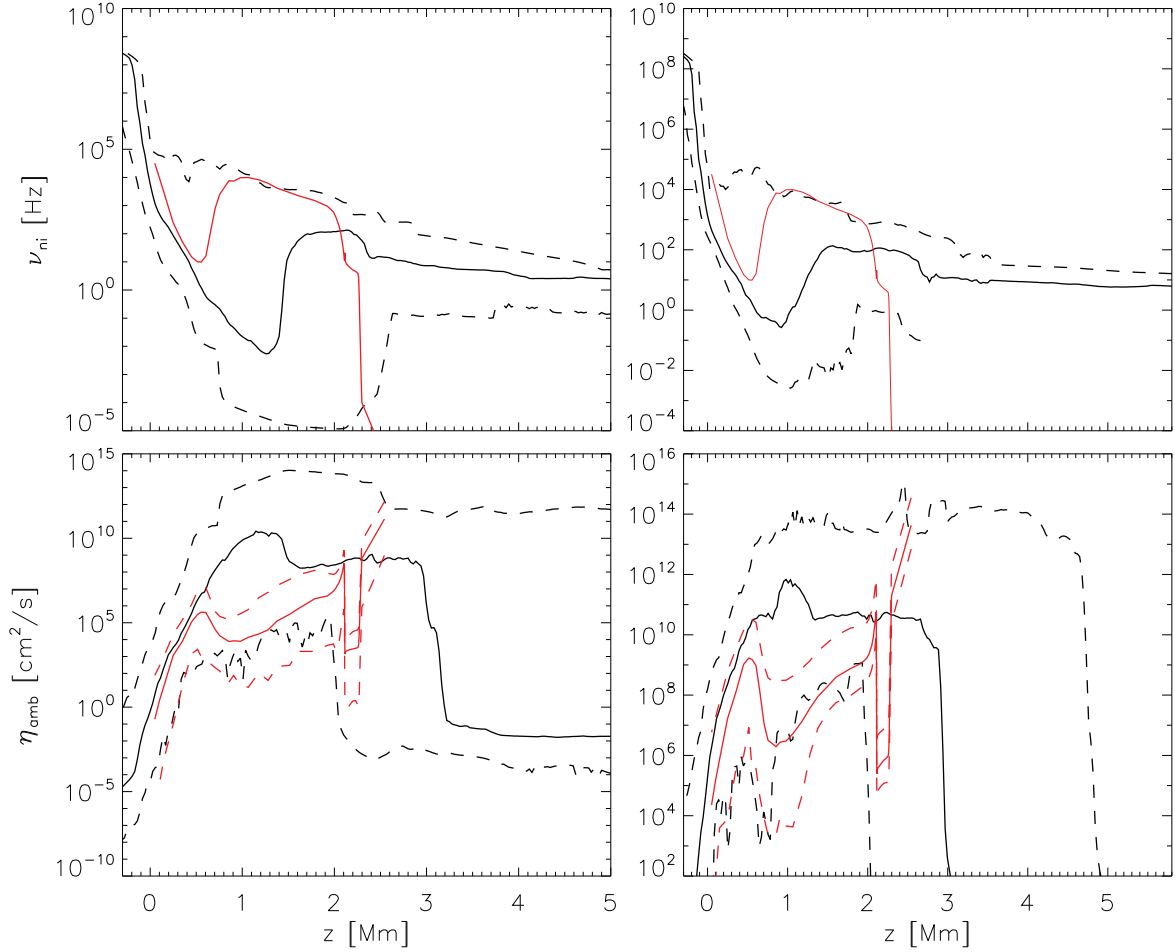}
 \caption{\label{fig:2dvsval} Minimum (dashed line), median (solid line) and maximum (dashed line) 
 of $\nu_{in}$ (top panels) and $\eta_{amb}$ (bottom panels) as function of height are shown for 
 the simulation labeled WB (black in left panels), SB (black in right panels)
 and for the VAL-C atmosphere (red). The VAL-C ambipolar 
 diffusion is calculated taking into account the maximum, minimum and median magnetic 
 field of the 2D models as a function of height. The minimum, median and maximum 
 are calculated in horizontal planes for the instant $t=500$~s.}
\end{figure}

\begin{figure}
  \includegraphics[width=0.95\textwidth]{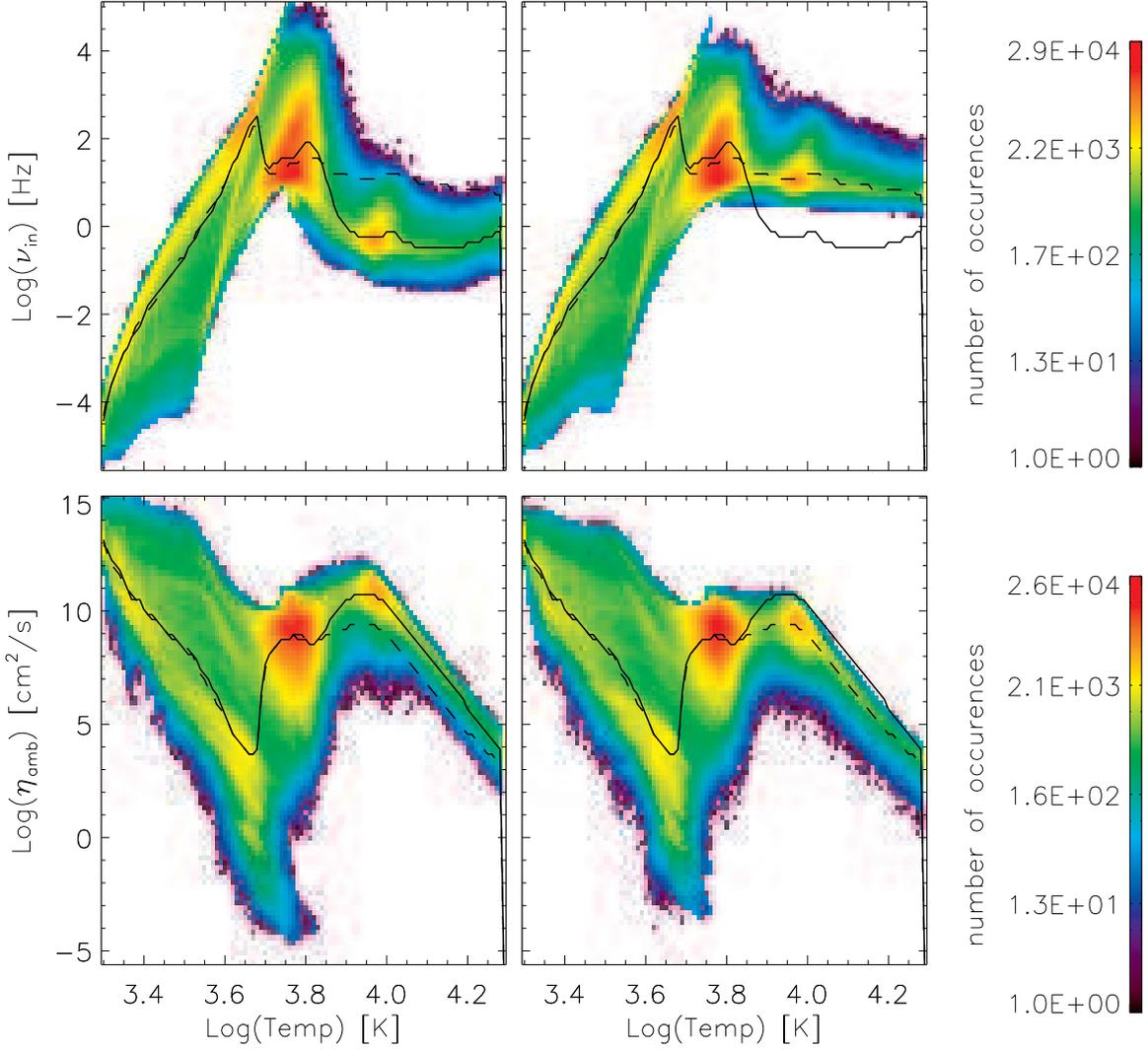}
 \caption{\label{fig:ABCw} Joint probability distribution function (JPDF) of the 
 temperature against of the ion-neutral collision frequency (top) and ambipolar 
 diffusivity (bottom) for the simulations labeled WB (left) and WC 
 (right). JPDF is calculated integrated over 220~s above the photosphere. 
 The colorbar is in logarithmic scale. The median as a function of temperature
 for the WB and WC cases are shown in solid, and dashed lines
 respectively.}
\end{figure}

\clearpage

\begin{figure}
  \includegraphics[width=0.95\textwidth]{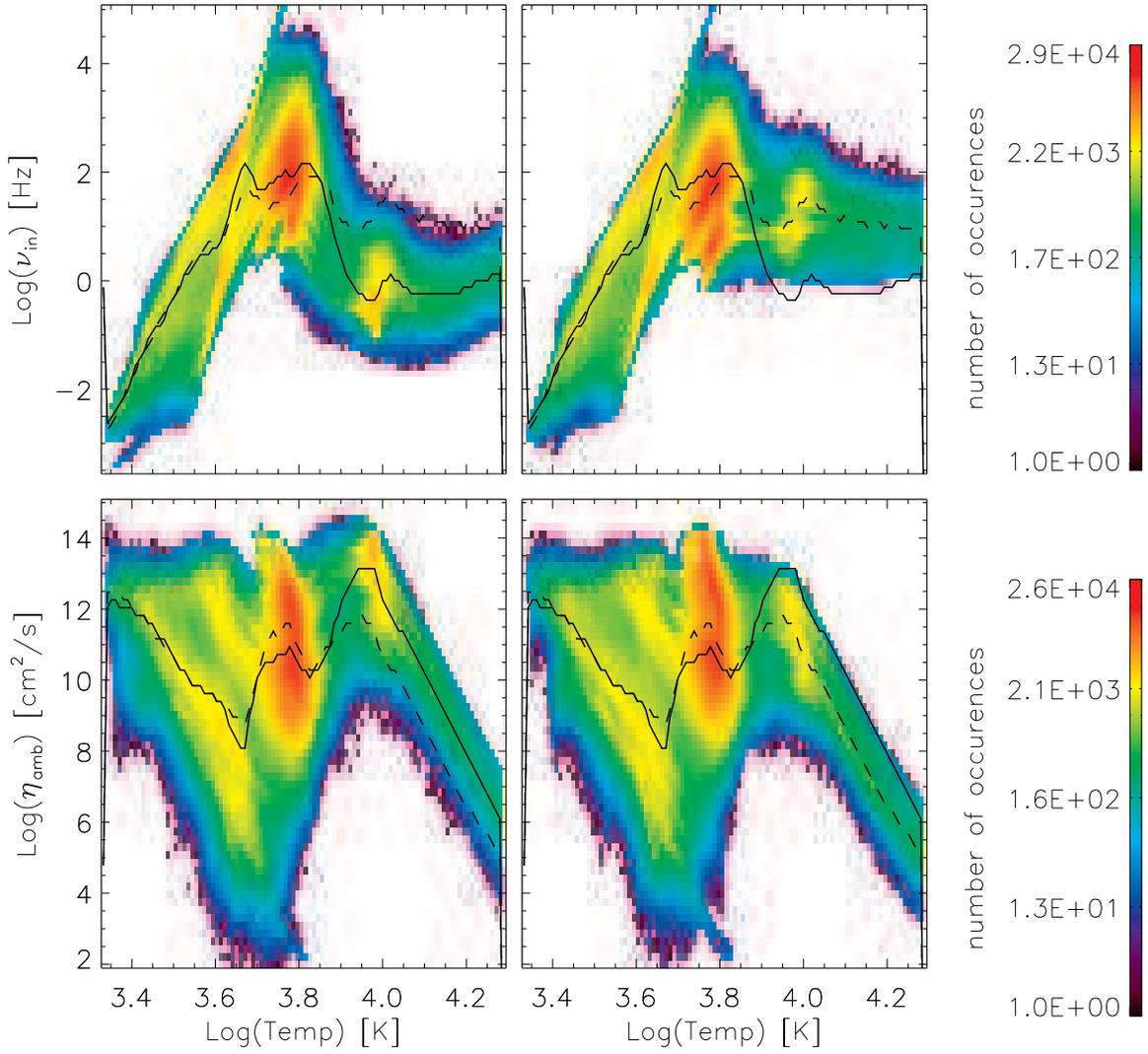}
 \caption{\label{fig:ABCs} The layout is the same as Figure~\ref{fig:ABCw}. However, 
 the simulations are SB (left) and SC (right).}
\end{figure}

\begin{figure}
  \includegraphics[width=0.95\textwidth]{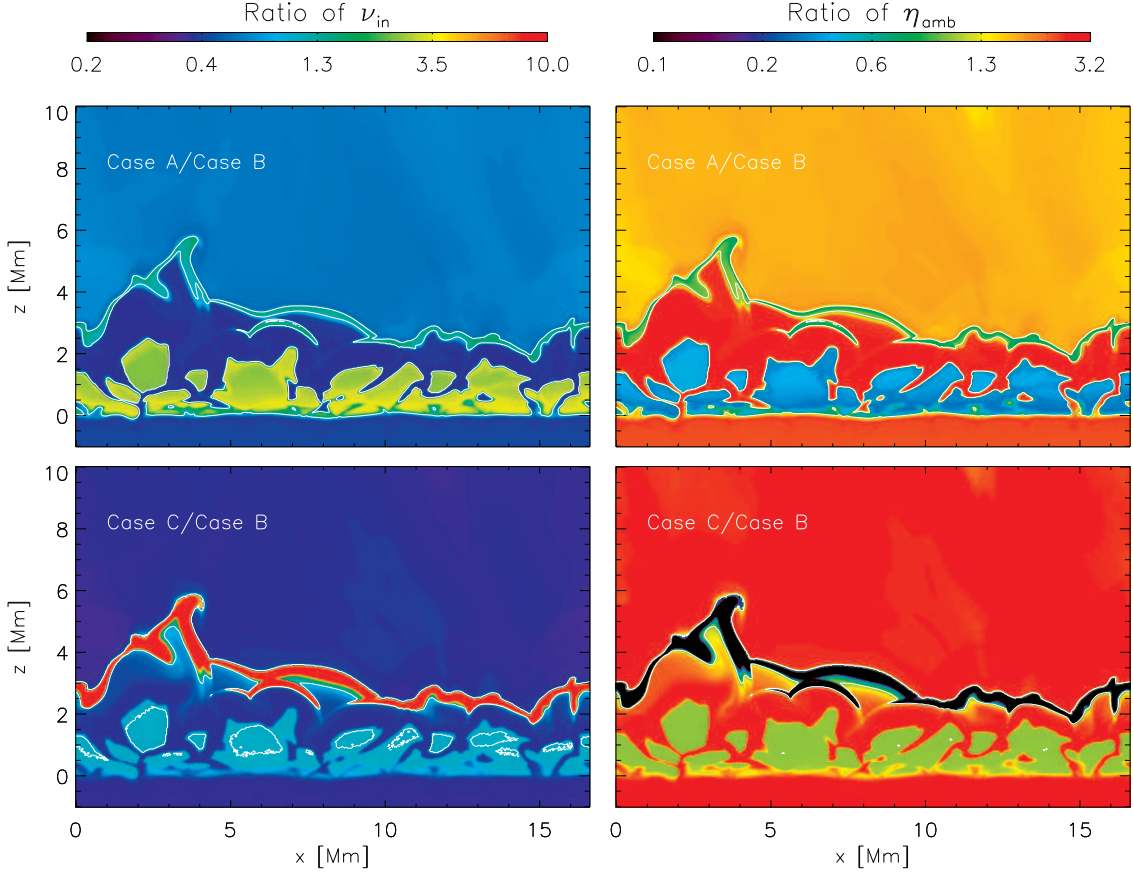}
 \caption{\label{fig:ABCdiffw} Ratio of the ion-neutral collision frequency (left panels) and
 ambipolar diffusivity (right panels) between case A to case B (top panels) 
 and case C to case B (bottom panels), for the atmosphere with weak magnetic field 
 strength. The white contours show where these ratios are equal to one. Note that the color
 scheme is in a logarithmic scale. We used the same atmospheric model for all three 
 cases, i.e., before the simulations diverge with time, but then calculated 
 the collision frequency and ambipolar diffusion using the different
 formulas of each case.}
\end{figure}

\begin{figure}
  \includegraphics[width=0.95\textwidth]{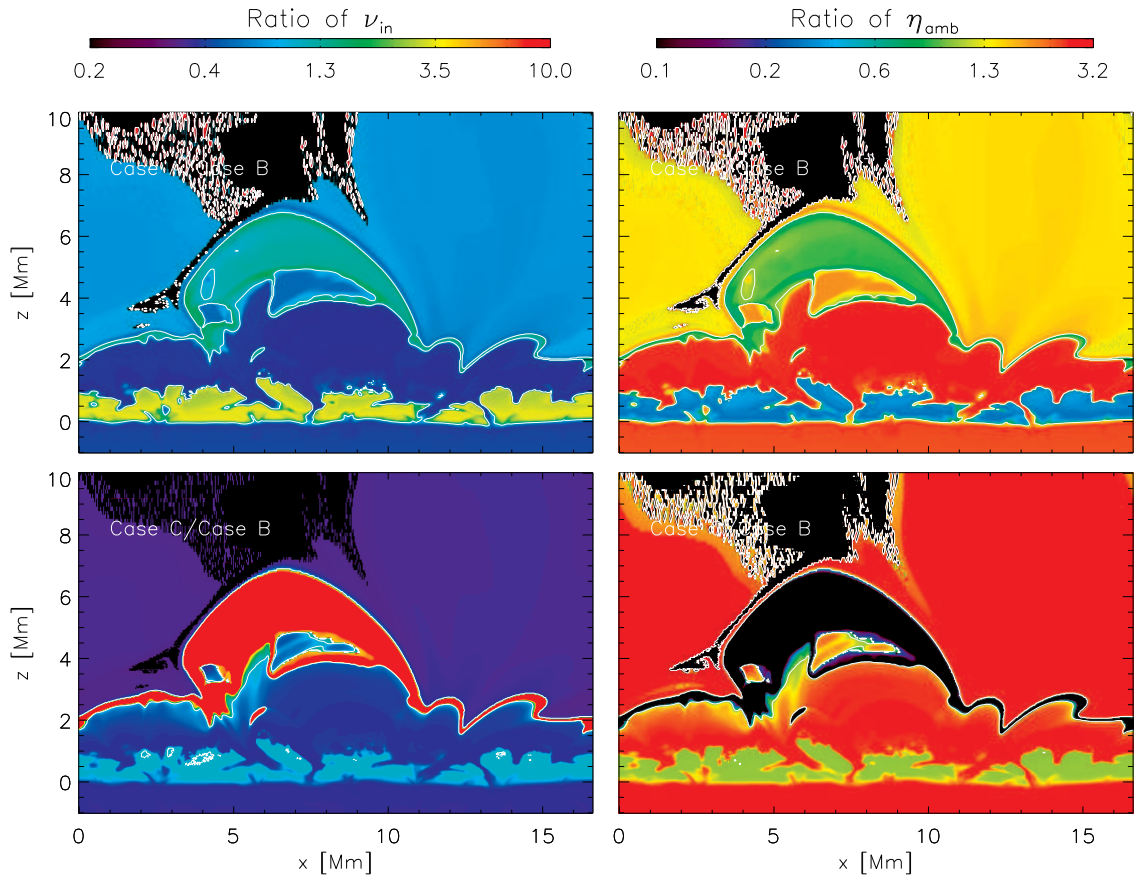}
 \caption{\label{fig:ABCdiffs} Same as Figure~\ref{fig:ABCdiffw} for the atmosphere with 
 strong magnetic field.}
\end{figure}

\begin{figure}
  \includegraphics[width=0.48\textwidth]{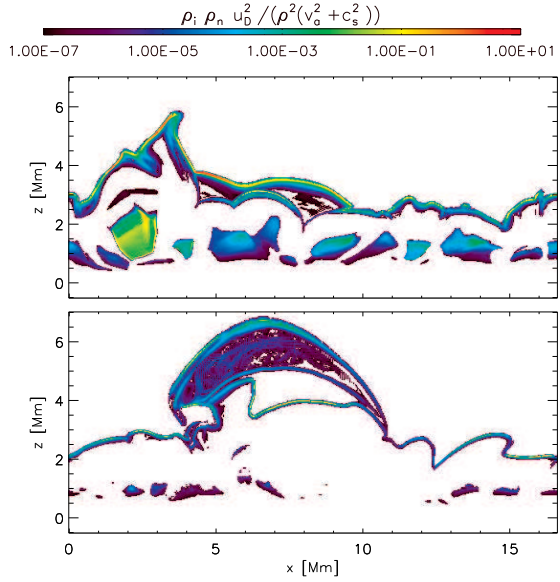}
 \caption{\label{fig:udvacs} The drift momentum has to be smaller
 than the fast momentum (Equation~\ref{eq:neg1}). The ratio between both
 terms, i.e.,  $\rho_i \rho_n {\bf u_D^2}$ and $\rho^2 (v_a^2+c_s^2)$ is 
 shown for the simulations labeled WB (top panel) and SB (bottom panel) 
 at $t=500$~s. The colorbar is in logarithmic scale and it is the same for 
 both panels.}
\end{figure}

\begin{figure}
  \includegraphics[width=0.48\textwidth]{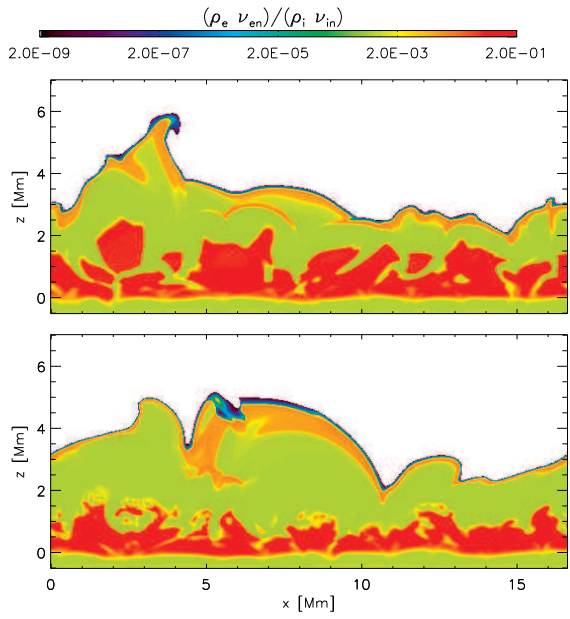}
 \caption{\label{fig:rcoll} Following the Equation~\ref{eq:neg2},  the ratio between 
 $\rho_e \nu_{en}$ and $\rho_i \nu_{in}$ is shown for the simulations labeled 
 WB (top panel) and SB (bottom panel) at $t=500$~s. 
  The colorbar is the same for both panels and it is in logarithmic scale.}
\end{figure}

\begin{figure}
  \includegraphics[width=0.95\textwidth]{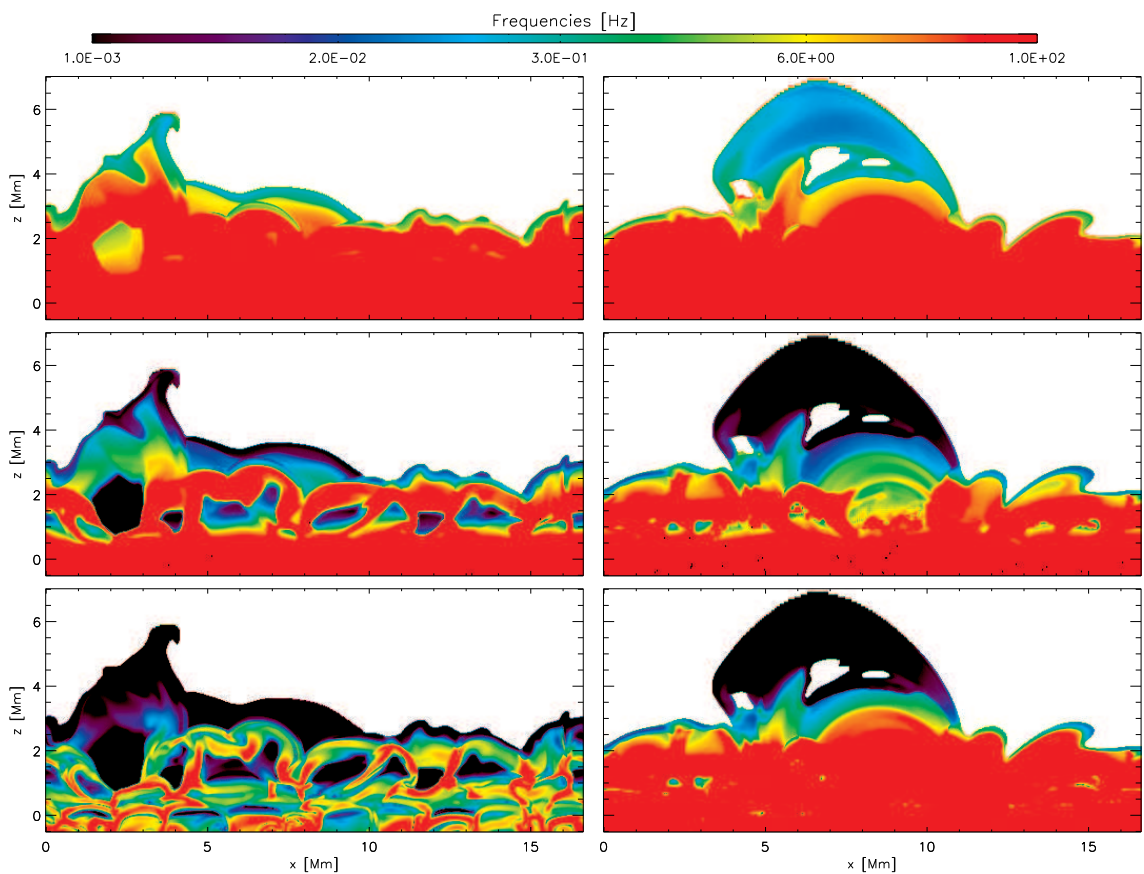}
 \caption{\label{fig:omegas} A study of the validity of the
   assumptions underlying the generalized Ohm's law. The dynamic
   frequency of the simulations ($\approx 0.5$ Hz) should remain lower
   than the frequency limits shown in the different panels for simulations 
 WB (left panels) and SB (right panels) at $t=500~s$. The frequencies are following the expressions Equation~\ref{eq:neg4} 
 (top panels), Equation~\ref{eq:neg5} (middle panels) and  Equation~\ref{eq:neg6} 
 (bottom panels). The colorbar for each frequency is located at the top side and 
 is in logarithmic scales. The white color is where the temperature is above $3\,10^{4}$K. } 
\end{figure}

\begin{deluxetable}{ccc}
\tablecaption{\label{tab:runs} Simulation description}
\tablehead{
\colhead{Name} & \colhead{Collision frequency}  & \colhead{Min/Mean/Max $|B|$ [G]} }
\startdata
WA & Case A &  0.003/0.25/3 \\
WB & Case B &  0.003/0.25/3 \\
WC & Case C & 0.003/0.25/3 \\
SA & Case A & 0.1/90/920 \\
SB & Case B & 0.1/90/920 \\
SC & Case C & 0.1/90/920 \\
\enddata
\tablecomments{ The left column lists the names of the different 2D simulations, middle column lists
the method used to calculate the collision frequency between ions and neutrals. 
The last column shows the minimum, mean and maximum value of the unsigned 
magnetic field strength in the photosphere.}
\end{deluxetable}

\begin{deluxetable}{ccccccccccccc}
\tablecaption{\label{tab:ions} Atomic info}
\tablehead{name &\vline& H &He & C & N & O & Ne & Na & Mg}
\startdata
abund &\vline&  12. &  11. &  8.55 & 7.93 &  8.77 & 8.51 &  6.18 & 7.48 \\
mass ion &\vline& 1.008 & 4.003 & 12.01 & 14.01 & 16. & 20.18 & 23. & 24.32 \\
$X_i$ &\vline&13.595 &  24.58 & 11.256 & 14.529 & 13.614 & 21.559 & 5.138 & 7.644  \\ \hline\hline
name &\vline& Al & Si & S & K & Ca & Cr & Fe &  Ni \\ \hline
abund &\vline& 6.4 &  7.55 &  7.21 &  5.05 &  6.33 &  5.47 &  7.5 &  5.08 \\
mass ion&\vline& 26.97 & 28.06 & 32.06 & 39.1 & 40.08 & 52.01 & 55.85 & 58.69 \\
$X_i$ &\vline& 5.984 & 8.149 & 10.357 & 4.339 & 6.111 & 6.763 & 7.896 & 7.633 \\ 
\enddata
\tablecomments{ The atomic species, abundances ($\log$ of number of atoms 
per $10^{12}$ Hydrogen atoms), mass ion (uma), and ionization fraction (eV)
of the 16 most important atomic species in the solar atmosphere are listed from 
the top to the bottom row. The various collision frequencies and the electron density are calculated taking
into account the atomic species in this table. }
\end{deluxetable}

\end{document}